\documentclass[aps,pra,twocolumn,preprintnumbers,amsmath,amssymb,showpacs,superscriptaddress]{revtex4}
\usepackage{bbm}
\usepackage{mathrsfs}
\usepackage{amsmath}
\usepackage[dvips]{graphicx}
\usepackage{epsfig}
\usepackage{color}

\usepackage[utf8]{inputenc}

\newcommand {\ud}  {\mathrm{d}}
        
\renewcommand{\*}{\cdot}
\renewcommand{\/}[2]{\frac{#1}{#2}}
\newcommand{\bra}{\langle}       \newcommand{\ket}{\rangle}
\renewcommand{\vec}[1]{\boldsymbol{#1}}
\newcommand{\mc}{\mathcal}
\newcommand{\pdag}{{\phantom{\dag}}}

\newcommand{\Span}{\operatorname{span}}

\newcommand{\eff}{\text{eff}}
\newcommand{\full}{\text{full}}
\newcommand{\orig}{\text{orig}}
\newcommand{\ua}{\uparrow}
\newcommand{\da}{\downarrow}
\newcommand{\SW}{\hat{\mc S}}
\renewcommand{\t}{\mathfrak{t}}
\newcommand{\veps}{\varepsilon}
\newcommand{\vphi}{\varphi}

\newcommand{\kzD}{|0,\ua\da\ket}
\newcommand{\bzD}{\bra 0,\ua\da|}
\newcommand{\kDz}{|\ua\da,0\ket}
\newcommand{\bDz}{\bra \ua\da,0|}
\newcommand{\kud}{|\ua,\da\ket}
\newcommand{\bud}{\bra \ua,\da|}

\newcommand{\bdu}{\bra \da,\ua|}

\newcommand{\aachen}{Institute for Theoretical Physics C, RWTH Aachen University, 52056 Aachen, Germany}
\newcommand{\munich}{Physics Department and Arnold Sommerfeld Center for Theoretical Physics, Ludwig-Maximilians-Universität München, 80333 München, Germany}
\newcommand{\queensland}{School of Physical Sciences, The University of Queensland, Brisbane, QLD 4072, Australia}

\begin{document}
\title{Magnetism, coherent many-particle dynamics, and relaxation with ultracold bosons in optical superlattices}

\author{T. Barthel} 
\affiliation{\aachen}
\author{C. Kasztelan}
\affiliation{\aachen}
\author{I. P. McCulloch}
\affiliation{\queensland}
\author{U. Schollw\"ock}
\affiliation{\munich}

\begin{abstract}
We study how well magnetic models can be implemented with ultracold bosonic atoms of two different hyperfine states in an optical superlattice. The system is captured by a two-species Bose-Hubbard model, but realizes in a certain parameter regime actually the physics of a spin-1/2 Heisenberg magnet, describing the second order hopping processes. Tuning of the superlattice allows for controlling the effect of fast first order processes versus the slower second order ones.
 Using the density-matrix renormalization-group method, we provide the evolution of typical experimentally available observables. The validity of the description via the Heisenberg model, depending on the parameters of the Hubbard model, is studied numerically and analytically.
The analysis is also motivated by recent experiments [S.\ F\"olling \emph{et al.}, Nature {\bf 448},  1029 (2007); S. Trotzky \emph{et al.}, Sience {\bf 319},  295 (2008)] where coherent two-particle dynamics with ultracold bosonic atoms in isolated double wells were realized. We provide theoretical background for the next step, the observation of coherent many-particle dynamics after coupling the double wells.
Contrary to the case of isolated double wells, relaxation of local observables can be observed.
The tunability between the Bose-Hubbard model and the Heisenberg model in this setup could be used to study experimentally the differences in equilibration processes for nonintegrable and Bethe ansatz integrable models. We show that the relaxation in the Heisenberg model is connected to a phase averaging effect, which is in contrast to the typical scattering driven thermalization in nonintegrable models. We discuss the preparation of magnetic groundstates by adiabatic tuning of the superlattice parameters.
\end{abstract}

\date{September 30, 2008}

\maketitle

\section{Introduction}\vspace{-0.5em}
One of the most exciting recent events in physics has been the increasing overlap between two previously disjoint fields, quantum optics and condensed matter physics. This has become possible due to the enormous progress in cooling dilute bosonic and also fermionic gases down to temperatures where respectively Bose-Einstein condensation and Fermi degeneracy (temperatures well below the Fermi energy) are reached.

A very attractive feature of this new class of experiments is that they provide the arguably cleanest realization of the (bosonic) Hubbard model \cite{Jaksch1998-81}, which with nearest-neighbor hopping and onsite interaction is the minimal model of strong correlation physics \cite{Hubbard1963-276}.

Here, we describe and analyze numerically a particular setup with ultracold bosons of two species in an optical superlattice, described by a Bose-Hubbard model. In the limit of strong onsite interactions, the system can be described by the spin-1/2 Heisenberg antiferro- or ferromagnet, depending on the parameters of the superlattice. The motivation is fourfold:

(i) In the vein of Feynman's idea to simulate quantum systems by other quantum systems \cite{Feynman1982-21}, it would be a great achievement to implement models of magnets like the Heisenberg model with ultracold atoms in optical lattices. In condensed matter systems, collective magnetism arises from the Coulomb interaction and the particle statistics which cause
 (super)exchange processes \cite{Heisenberg1926-39,Dirac1926-112,Heisenberg1928-49,Dirac1929-123,Kramers1934-1,Anderson1950-79,Anderson1959-115}.
In particular, exchange
 interactions resulting from second order hopping processes in the Fermi-Hubbard model dominate its behavior in the limit of strong onsite interaction and are captured by an effective spin model, namely the Heisenberg antiferromagnet \cite{Anderson1959-115,Chao1977-10,Fazekas1999}.

While collective magnetism has been widely studied in solids over the decades, several experimental restrictions apply quite generally: It is generally far from clear to what extent the typical simplified models are quantitatively realistic, and how to obtain the interaction parameters. External control of these parameters is very difficult. Moreover, quantum magnetism becomes particularly interesting in low dimensions. In real effectively low-dimensional solids it is however hard to control or assert the effect of the weaker interactions in the second and/or third dimension. Last but not least, solids give us only access to the linear response regime as sampled e.g.\ by neutron scattering. Questions of out-of-equilibrium many-body dynamics are essentially inaccessible.

Experiments with ultracold atoms in optical lattices constitute clean and well-tunable manifestations of the Bose- or Fermi-Hubbard model \cite{Jaksch1998-81,Bloch2007}. To implement magnetic systems, the most straight forward approach would hence be to use a gas of ultracold fermions. However, cooling of fermionic gases to the quantum regime is a considerably harder task due to the lack of s-wave scattering among identical fermions  \cite{DeMarco1999-285,DeMarco1999-92,OHara2002-298,Koehl2005-94,Chin2006-443}. Hence it is desirable to develop an alternative route via gases of ultracold bosons. Our investigations follow this idea \cite{Kuklov2003-90,Duan2003-91,Altman2003-5,Garcia-Ripoll2004-93,Barmettler2008-78}.
Although we focus here on one dimension, analogous setups in higher dimensions \cite{Sebby-Strabley2006-73} could be used to investigate a plethora of frustrated spin systems that are hard to access analytically and numerically.

(ii) The superlattice structure chosen in our setup (in analogy to the recent experiments \cite{Foelling2007,Trotzky2008-319}) allows in contrast to \cite{Duan2003-91,Altman2003-5} for the tuning of the effective spin-spin interaction by changing an alternating scalar potential $\Delta$; Fig~\ref{fig:systemParameters}. With the hopping strength $\t$ and the onsite interaction $U$ of the Hubbard model, the coupling in the corresponding effective spin model is then ${4\t^2 U}/(U^2-\Delta^2)$. This allows on the one hand to switch for the effective model between the Heisenberg ferro- and antiferromagnet.  On the other hand one might hope to increase for a fixed onsite interaction $U$ the effective coupling, by choosing $\Delta\sim U$. In this case, the relevant physics would become visible at correspondingly higher temperatures. However, the validity of the effective model breaks down in the vicinity of $\Delta\sim U$. So one has to balance the validity of the Heisenberg description and the temperatures needed to observe the quantum effects. To this purpose, the parameter $\Delta$ can be easily varied and used to tune to the Heisenberg regime in controlled fashion.

(iii) In recent experiments \cite{Foelling2007,Trotzky2008-319} by the Bloch group, the same optical superlattice as the one discussed here was used. But its parameters were chosen such that the superlattice decomposed actually into isolated double wells. The experiments analyzed dynamics in these double wells and contrasted in particular first order (hopping) processes (Hubbard regime) versus slower second order processes (Heisenberg regime). The next step would be to observe coherent many-particle dynamics after coupling the double wells. We analyze such a situation by the time-dependent density matrix renormalization group method (DMRG) \cite{Daley2004,White2004}. We focus on the coherent evolution of an initial Néel state and the differences between the Heisenberg and the Hubbard regimes and present the experimentally available observables.

(iv) Besides testing the coherence in the experiments, the setup allows to address questions of non-equilibrium many-particle systems, which is in general difficult for all present analytical and numerical methods. Contrary to the setup of isolated double wells, one observes for the many-particle dynamics in our setup a relaxation of local quantities. This is an indicator for convergence of subsystems with finite real-space extent to a steady state.  Recently, the mechanism of how such a relaxation may occur was clarified for (free) integrable systems in \cite{Barthel2008-100}. Corresponding examples can also be found in \cite{Rigol2007-98,Cazalilla2006-97,Cramer2007,Gangardt2007}. For a few nonintegrable systems the question was analyzed numerically in \cite{Kollath2007-98,Manmana2007-98,Cramer2008-101} and analytically e.g.\ in \cite{Moeckel2008-100,Eckstein2007}. In general one expects that in nonintegrable models, thermalization occurs (due to scattering processes), and that in integrable models, relaxation (to a nonthermal steady state) occurs via phase averaging effects \cite{Barthel2008-100}. This is demonstrated here analytically for the Heisenberg model.
Our setup could be used to study experimentally such relaxation processes -- in particular, the qualitative differences between nonintegrable systems, here the Bose-Hubbard model, and Bethe ansatz integrable models \cite{Bethe1931,Zachary1996}, here the Heisenberg model.

We also fill a certain gap of current literature on such topics (see e.g.\ \cite{Duan2003-91,Altman2003-5,Garcia-Ripoll2004-93}) by emphasizing that the Heisenberg spins of the effective model, obtained by the Schrieffer-Wolff transformation (in Appendix~\ref{sec:derivation}) \cite{Schrieffer1966-149}, should not be identified directly with the two boson species. A spin up of the effective model corresponds rather to a particle of species 1 dressed by hole-double-occupancy fluctuations. The analogy holds only for small $\t/(U\pm\Delta)$. The consequences for experimentally available observables are surprisingly strong.
Recently in \cite{Barmettler2008-78} a setup of coupled double wells ($\Delta=0$, but alternating hopping $\t\neq \t'$ in Fig.~\ref{fig:systemParameters}) was analyzed numerically -- in particular, the possibilities to generate entangled pairs of particles were studied. Again, a perfect mapping to a spin model was assumed from the outset. The validity of this mapping is one central topic of our article.

The paper is organized as follows. Section~\ref{sec:setupAndModel} describes the experimental setup and how it can be described by a Bose-Hubbard model. Restricting to half filling, Section~\ref{sec:effectiveModel} and Appendix~\ref{sec:derivation} derive by a Schrieffer-Wolff transformation for the limit of large onsite interactions an effective model which is the Heisenberg antiferro- or ferromagnet. In Section~\ref{sec:Evolution}, we investigate by time-dependent density-matrix renormalization-group (DMRG) the evolution of typical observables like magnetization, momentum-space and real-space correlators, where the first two are also available experimentally. The focus is on contrasting the differences between the full Hubbard dynamics and the corresponding effective spin model, and also the differences to the case of isolated double wells \cite{Foelling2007,Trotzky2008-319}. We observe indications for (local) relaxation to steady states. This is discussed in Section~\ref{sec:relaxation} where we also explain how the relaxation for the Heisenberg model is connected to a phase averaging effect. Section~\ref{sec:validity} addresses in more detail the question why and under what circumstances the effective model is a valid description for the full Hubbard Hamiltonian, especially for the dynamics. In Section~\ref{sec:groundstates} we argue that the groundstate of the Heisenberg antiferromagnet could be prepared by tuning an alternating hopping parameter of the superlattice adiabatically. Section~\ref{sec:conclusion} gives a short conclusion.

\section{Setup and model} \label{sec:setupAndModel}
\begin{figure}[ht]
\epsfig{file=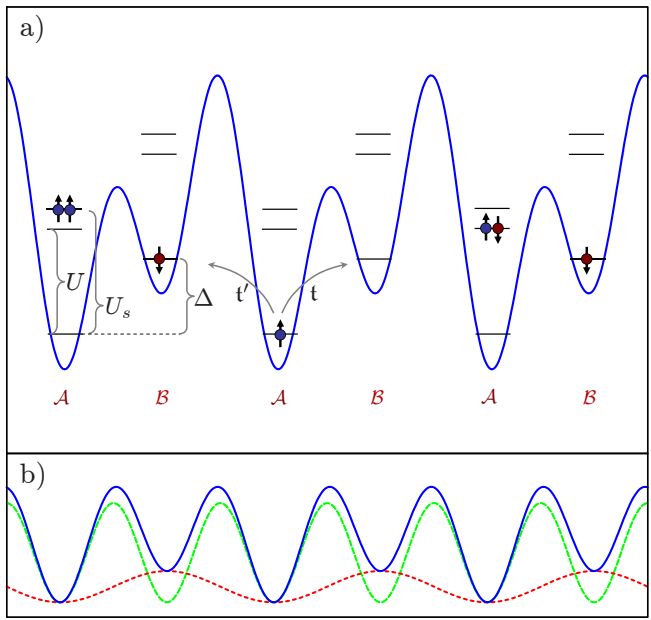, angle=0,width=1\linewidth}
\caption{a) Tight-binding system parameters for our system of ultracold bosons of two species ($\ua$, $\da$) in a one-dimensional optical superlattice \eqref{eq:Ham-Hubbard}. b) If the phase shift between the two laser potentials (dashed lines) is zero, the minima of the full potential (solid line) are equidistant, the Wannier wave functions for each site are reflection symmetric, and hopping parameters $\t$ and $\t'$ are hence equal.}
\label{fig:systemParameters}
\end{figure}
In the following, we present a setup of ultracold bosonic atoms in a one-dimensional optical superlattice that reduces in certain parameter regimes, where first order hopping processes are suppressed, to the Heisenberg ferro- or antiferromagnet. The use of bosons is motivated by the fact that experimentally, access to the low energy quantum physics is at the moment still much harder for fermionic systems.
In analogy to the fermionic case, for which the antiferromagnetic Heisenberg model describes the effective low-energy physics of the fermionic Hubbard model, we choose to have two species $\sigma\in\{\ua,\da\}$ of bosons in the lattice -- two hyperfine states of a bosonic atom.  At half filling ($N$ sites, $N_\ua=N_\da=N/2$), the effective low energy model is, as we will see later, the ferromagnetic Heisenberg model. To allow for tuning the effect of first order processes and to switch between a ferromagnetic and antiferromagnetic regime, we employ an alternating onsite potential $\Delta_i$ and call the two sublattices $\mc A$ and $\mc B$. The potential minima differ by a value $\Delta>0$. Such superlattices can be generated by the superposition of two laser frequencies of ratio $1:2$, see Ref.~\cite{Anderlini2006-39,Foelling2007,Trotzky2008-319} and Fig.~\ref{fig:systemParameters}. Further, the tight-binding approximation with restriction to the first Bloch band (one Wannier function per site) is assumed.
Then the system is described by the two-species Bose-Hubbard Hamiltonian
\begin{multline}
\label{eq:Ham-Hubbard}
\hat H=-\t\sum_{\sigma,\bra ij\ket}(a_{\sigma i}^\dag a_{\sigma j}^\pdag + h.c.)
 + \sum_{\sigma,i} \Delta_i n_{\sigma,i}\\
 + U\sum_i n_{\ua i}n_{\da i}
 + \/{U_s}{2} \sum_{\sigma,i} n_{\sigma i}(n_{\sigma i}-1),
\end{multline}
In particular, we choose 
\begin{equation}
\label{eq:anisotropy}
\Delta_i
=\begin{cases}
-\Delta/2& \text{for even } i\\
\Delta/2& \text{for odd } i.
 \end{cases}
\end{equation}
The superlattice potential is of the form
\begin{equation}\label{eq:laserPotential_x}
 V(x)=V_0 \sin^2(kx)+V_1\sin^2(kx/2+\phi).
\end{equation}
 The amplitude $V_1$ of the second potential can be used to tune $\Delta$; for our purposes, $V_1\ll V_0$. In \eqref{eq:Ham-Hubbard} it was assumed that hopping from a site $i$ to its neighbors $i\pm 1$ occurs with equal amplitude $\t$. In principle, the hopping depends exponentially on the distance between the potential minima \cite{Jaksch1998-81,Zwerger2003-5}. Only if we choose the phase difference $\phi$ to vanish as in Fig.~\ref{fig:systemParameters}b, the positions of the potential minima will be equidistant for all $V_1$, and the Wannier wave functions for all sites are reflection symmetric. In this case, the hopping will hence be of equal strength for all bonds, i.e.\ $\t=\t'$  in Fig.~\ref{fig:systemParameters}. This situation is considered in the following.

For the actual analysis of dynamics in Section~\ref{sec:Evolution}, we will choose equal inter- and intra-species interaction, $U=U_s$. This is at the moment the standard situation for the corresponding experiments. 
With $U=U_s$, the effective Heisenberg models, describing the second order physics, will turn out to be isotropic.
The onsite potential $\Delta$ and interaction $U$ can be calculated in harmonic approximation of the lattice potential around its minima. For $\phi=0$ in \eqref{eq:laserPotential_x}, the corresponding oscillator frequencies are $\hbar\omega_\pm =2\sqrt{E_r (V_0\pm V_1/4)}\approx 2\sqrt{E_rV_0}(1\pm \/{1}{8}\/{V_1}{V_0})$, where $E_r\equiv \/{\hbar^2 k^2}{2m}$ is the recoil energy of the laser potential with the shorter wave length. For the dependence of the onsite potential on the lattice parameters follows
\begin{equation}\label{eq:Delta}
\textstyle
 \Delta = V_1-\/{\hbar}{2}(\omega_+-\omega_-)\approx V_1\left(1-\/{1}{4}\sqrt{\/{E_r}{V_0}}\right).
\end{equation}
In order to achieve an effectively one-dimensional lattice, $V(x)$ is superimposed with two transversal laser beam potentials of a (higher) amplitude $V_\bot$. Within the harmonic approximation, and with the s-wave scattering length $a_s$, the resulting onsite interactions for two neighboring sites are \cite{Zwerger2003-5}
\begin{equation}
 U_\pm=\sqrt{\/{8}{\pi}} a_s k \left(\/{(V_0\pm V_1 /4) V_\bot^2}{E_r^3}\right)^{\/{1}{4}} E_r.
\end{equation}
They are in principle not identical. Irrespective of this, the effective spin model derived in Section~\ref{sec:effectiveModel} would be isotropic and translation invariant. As we will see in the following for a set of realistic experimental parameters, we have for the case $V_1=0$ ($\Rightarrow$ $\Delta=0$) that $V_0\gg E_r$ and $V_0\gg U$. The onsite potentials $\Delta$ considered in our numerical simulations of the Hubbard model \eqref{eq:Ham-Hubbard} are from the interval $\Delta\in[0,4U]$.  According to \eqref{eq:Delta}, $V_1$ will for nonzero $\Delta$ hence obey $V_0\gg V_1$ and we can thus use $U_+=U_-\equiv U$ in good approximation.

For the tight-binding approximation to hold, we need that energies $\t$, $U$, and $\Delta$ are well below $\hbar\omega_\pm\approx 2\sqrt{E_r V_0}$, the energy scale for vibrations of an atom in one minimum of the laser potential. With the hopping $\t=\/{4}{\sqrt{\pi}}\left({V_0 V_\bot^2}/{E_r^3}\right)^{1/4}e^{-2\sqrt{{V_0}/{E_r}}}E_r$  \cite{Zwerger2003-5,Bloch2007}, we have for example with $\lambda={2\pi}/{k}=800nm$, Rubidium atoms (i.e.\ $a_s k\approx \pi 0.01$), $V_0\approx 8.7 E_r$ and $V_\bot\approx 30 E_r$ (cmp.\ e.g.\ to \cite{Foelling2007}) that $\t\approx 0.06 E_r$, $U\approx 8\t$ and $\hbar\omega_0\approx 100\t$. So as long as $\Delta$ is also well below $ 100\t$, the tight-binding approximation for \eqref{eq:Ham-Hubbard} using the lowest Bloch band is valid. This will be the case in the rest of the paper.

\section{Effective model} \label{sec:effectiveModel}
\begin{figure}[ht]
\epsfig{file=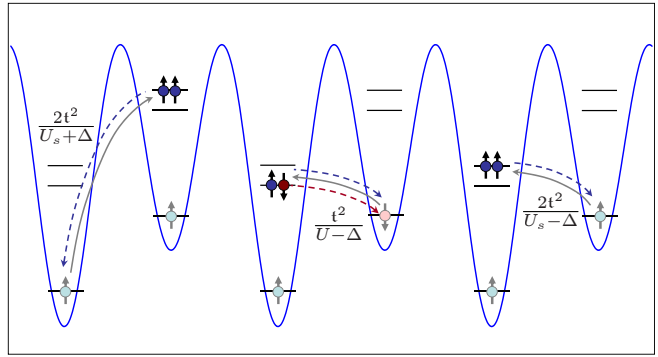, angle=0,width=1\linewidth}
\caption{Some second order hopping processes (superexchange) contributing to the effective spin model \eqref{eq:HamEffH1spinA} when first order hopping processes in the full Hubbard model \eqref{eq:Ham-Hubbard} are suppressed. The figure displays possible initial states (light color) with one particle per site and intermediate states (dark color) with doubly occupied and empty sites.}
\label{fig:superExchange}
\end{figure}

To go to a regime where the physics of the two-species Bose-Hubbard model \eqref{eq:Ham-Hubbard} reduces to that of a Heisenberg magnet, we choose half filling
\begin{equation}
\label{eq:half-filling}
N_\da=N_\ua=N/2,
\end{equation}
($N$ is the total number of lattice sites) and assume the
 large-$U$ limit
\begin{equation}
\label{eq:largeU-limit}
\t\ll |U\pm\Delta|.
\end{equation}
In this limit, occupation of a single site by more than one boson is energetically unfavorable and will occur only in short lived intermediate states. This means that single (first order) hopping processes are suppressed. Besides some hybridization effects, exactly one boson sits on each lattice site and can be identified with an effective spin on that site (up and down orientations corresponding each to one of the boson species).
The second order hopping processes as depicted in Fig.~\ref{fig:superExchange} lead then to nearest neighbor spin-spin interactions.

We will show below, how the effective Hamiltonian can be derived by a Schrieffer-Wolff transformation. While this is a well-known procedure, there is an interesting twist to the interpretation of the result. For the moment let us work with the naive identification of spins up and down of the effective model with the two boson species of the full model \eqref{eq:Ham-Hubbard}.
We want to derive an effective Hamiltonian describing the physics of the Hubbard Hamiltonian \eqref{eq:Ham-Hubbard} in the subspace $\mc H_1$ of singly-occupied sites.
\begin{equation} \label{eq:subspaceH1}
\mc H_1:=\Span\{|\ua\ket,|\da\ket\}^{\otimes N}
\end{equation}
The effective Hamiltonian can be deduced from the following simple recipe: With exactly one spin per site, the onsite interaction is ineffective. Hopping processes occur only in second order, leading to a spin-spin interaction
\begin{multline}
\label{eq:HamEffH1spinA}
\hat H_\eff= 
- J \sum_{\bra ij\ket} (\hat S^x_i\hat S^x_{j}+\hat S^y_i\hat S^y_{j})
+ (J-J_{s}) \sum_{\bra ij\ket} \hat S^z_i\hat S^z_{j}.
\end{multline}
The corresponding coupling strengths $J$ and $J_s$ are obtained by dividing for each possible second order process (Fig.~\ref{fig:superExchange}) the product of the transition matrix elements $\t\*\t$ for the hopping to a neighboring site and back by the energy difference $U\pm \Delta$ ($U_s\pm \Delta$) to the intermediate state, and adding all such terms that contribute to the same effective spin-spin interaction (see also \cite{Trotzky2008-319}).
\begin{equation}
\label{eq:HamEffH1spinCouplings}
J 
= \/{2 \t^2}{U+\Delta}+\/{2 \t^2}{U-\Delta}
=\/{4\t^2U}{U^2-\Delta^2},\,\,\,
J_{s}
=2\/{4\t^2U_s}{U_s^2-\Delta^2}.
\end{equation}
The effective Hamiltonian \eqref{eq:HamEffH1spinA} is the XXZ model. From now on we specialize to $U=U_s$, i.e.\ $J-J_s=-J$ and have hence the isotropic Heisenberg ferromagnet for $\Delta<U$ ($J>0$) and the isotropic antiferromagnet for $\Delta>U$ ($J<0$). Note that the effective Heisenberg Hamiltonian would also be isotropic and translation invariant if the onsite interaction $U$ would be different for even and odd sites. If the hopping would be alternating ($\t\neq\t'$), we would obtain the dimerized Heisenberg model.

A common mathematical approach for the deduction of such effective models is the Schrieffer-Wolff transformation \cite{Schrieffer1966-149}. We are interested in the physics of the subspace $\mc H_1$ with exactly one particle per site. The Hubbard Hamiltonian couples this subspace in first order of the hopping $\t$ to the rest of the Hilbert space (states with doubly occupied and empty sites). The Schrieffer-Wolff transformation,
\begin{equation} \label{eq:Schrieffer-Wolff}
\hat H_{\eff}^\full:=e^{i\SW} \hat H e^{-i\SW},\quad
 \mc H_1^\orig := e^{-i\SW}\mc H_1,
\end{equation}
is a unitary transformation with generator $\SW$ chosen such that the transformed Hamiltonian $\hat H_{\eff}^\full$ does not contain terms anymore that couple $\mc H_1$ to the rest of the Hilbert space, or at least only in some higher order of $\t$. In Appendix~\ref{sec:derivation}, a generator
\begin{equation} \label{eq:SWtrafo-orderA}
\SW=\mc O\left(\/{\t}{U \pm \Delta}\right)
\end{equation}
is derived, such that effective Hamiltonian is
\begin{equation} \label{eq:HamEffH1spin}
\hat H_{\eff}=\hat H_{\eff}^\full|_{\mc H_1}
= - J \sum_{\bra ij\ket} \hat{\vec{S}}_i\*\hat{\vec{S}}_j
 + \mc O(\t^4).
\end{equation}
The full effective Hamiltonian $\hat H_{\eff}^\full$ \eqref{eq:HamEff} still contains a term $i[\SW,\hat H_\t^0]$ representing the remaining coupling of the subspace $\mc H_1$ to the rest of the Hilbert space which is of order $\t^2$.

The method is based on the smallness of $\SW$. According to \eqref{eq:SWtrafo-orderA}, it hence breaks down when $U\sim |\Delta|$. The effective Hamiltonian \eqref{eq:HamEffH1spin} is only valid for $|U\pm \Delta|\gg \t$. Only in this regime, the first order hopping processes leading out of $\mc H_1$ are suppressed. See the discussion in Section~\ref{sec:validity}.

Often spin up (down) states of the effective model are then identified with a boson of species $\ua$ (species $\da$) on the corresponding sites. However, with respect to the original model, Eq.~\eqref{eq:Ham-Hubbard}, it is \emph{not} $\mc H_1$ itself that is weakly coupled to the rest of the Hilbert space and evolves according to the Heisenberg Hamiltonian but the subspace $\mc H_1^\orig$ defined in Eq.~\eqref{eq:Schrieffer-Wolff}. A spin up in the effective model corresponds rather to a boson of species $\ua$ with a cloud of hole-double-occupancy fluctuations $a_{\sigma i}\to e^{i\SW} a_{\sigma i} e^{-i\SW}$; see also Fig.~\ref{fig:lattice_DoubleWell_SW-Trafo}. The experimental consequences are surprisingly strong, as we will in the next section.

\section{Time-evolution from the N\'eel state} \label{sec:Evolution}
In the following, numerical results for the evolution of the system where the initial state is the N\'eel state
\begin{equation}
\label{eq:NeelState}
|\phi\ket :=|\ua\da\ua\da\ua\da\ldots\ket\in \mc H_1
\end{equation}
are presented. This parallels recent experimental investigations \cite{Foelling2007,Trotzky2008-319} of the evolution of corresponding states $|\ua\da\ket$ in isolated double wells.

To compare the effect of first and second order processes, the evolution was done twice for each set of parameters ($\t=1$, $U=8$, various $\Delta$; see Section~\ref{sec:setupAndModel}), once with the corresponding Heisenberg model (in the subspace $\mc H_1$) and once with the full Hubbard Hamiltonian (in the full Hilbert space), where the initial state \eqref{eq:NeelState} was in fact chosen as the tensor product of alternatingly having one boson of species $\ua$ or $\da$ on each site. The two different time scales of first and second order processes become clearly visible. The qualitative differences to the isolated double well situation (as analyzed in \cite{Foelling2007,Trotzky2008-319}) and resulting interesting questions for experimental investigations are discussed.

\subsection{Errors through experimental limitations in state preparation and measurement}  \label{sec:errors}
We shortly want to discuss how well the dynamics of the magnetic model, the Heisenberg model, can be implemented experimentally by those of the two-species Bose-Hubbard model. In the literature on magnetism via ultracold two-species atom gases in optical lattices  \cite{Duan2003-91,Altman2003-5,Barmettler2008-78}, spins up and down of the magnetic system are usually identified directly with atoms of species $\ua$ and $\da$ of the ultracold gas. In this vein, evolution of the N\'eel state \eqref{eq:NeelState} with the Heisenberg Hamiltonian would be translated into evolution of the state $|\ua\da\ua\da\ua\da\dots\ket$ with the Hubbard Hamiltonian. This is actually correct only to zeroth order in $\SW$.

We want to implement the evolution of a state $|\phi\ket\in\mc H_1$, \eqref{eq:subspaceH1}, under the effective Hamiltonian $\hat H_\eff = e^{i\SW}\hat H e^{-i\SW}|_{\mc H_1}$ by the evolution of a state $|\psi\ket=e^{-i\SW}|\phi\ket\in\mc H_1^\orig$, \eqref{eq:Schrieffer-Wolff}, under the Bose-Hubbard Hamiltonian $\hat H$. The state $|\psi\ket$ is $|\phi\ket$, superimposed with states where starting from $|\phi\ket$, pairs of doubly occupied sites and empty sites were created (e.g.\ Eq.~\eqref{eq:2site-psi} and Fig.~\ref{fig:lattice_DoubleWell_SW-Trafo} below). This can also be interpreted as constructing the N\'eel state with effective spins, each corresponding to a boson accompanied by a cloud of hole-double-occupancy fluctuations $a_{\sigma i}\to e^{i\SW} a_{\sigma i} e^{-i\SW}$.
The decisive point is now that it seems not possible to prepare such states from $\mc H_1^\orig$ (and has to our knowledge never been done), but only some specific states from $\mc H_1$. Hence, instead of starting the experiment from the initial state $|\psi\ket$, one is forced to neglect the Schrieffer-Wolff transformation and start from the state $|\phi\ket$ -- in our example the N\'eel state. For observables that do not change the number of doubly occupied sites, this results in an error of $\mc O(\SW^2)$.

If we had determined the exact Schrieffer-Wolff transformation $e^{-i\SW}$ (i.e.\ $\SW$ exact to all orders in $\t$) and could actually implement it e.g.\ by time evolution in the experiment, all measurements would be exact. One could prepare the state $|\phi\ket$, apply the Schrieffer-Wolff transformation by time evolution to obtain $|\psi\ket$, evolve with the Hubbard Hamiltonian for some time $t$, apply the inverse Schrieffer-Wolff transformation and measure our observable $\hat O$. This would yield the exact equality
\begin{multline} \label{eq:measurementCorrect}
 \bra\phi|e^{-\hat H_\eff t/i\hbar}\*\hat O \* e^{\hat H_\eff t/i\hbar}|\phi\ket \\
= \bra\phi|e^{i\hat S}e^{-\hat Ht/i\hbar}e^{-i\hat S}\*\hat O \*e^{i\hat S}e^{\hat Ht/i\hbar}e^{-i\hat S}|\phi\ket, 
\end{multline}
where $\hat H_\eff$ would now of course be a generalization of the Heisenberg model with longer ranged interactions.

In the Appendix~\ref{sec:SWtrafo}, $\SW$ is determined to first order in $\/{\t}{U\pm\Delta}$, \eqref{eq:SWtrafo-order}, and correspondingly $\hat H_\eff$ to first order in the effective coupling $J$, \eqref{eq:HamEffH1spin2}. Using this approximation of $\SW$ instead of the exact one, the remaining errors in the observables are of order $\SW^4$, i.e.\ $\mc O((\/{\t}{U\pm\Delta})^4)$. (It is not $\mc O(\SW^2)$, because the operator $\SW$, given in Eq.~\eqref{eq:SWtrafo-result}, changes the number of double occupancies by one and the typical observables $\hat O$ we are interested in do not.)
However, failing to implement the Schrieffer-Wolff transformation completely, i.e.\ measuring $\bra\phi|e^{-\hat Ht/i\hbar}\*\hat O \*e^{\hat Ht/i\hbar}|\phi\ket$ instead of \eqref{eq:measurementCorrect}, leads to errors of order $\SW^2$. This will be demonstrated in an example (Section~\ref{sec:magnetization}, Figs.~\ref{fig:twoSiteEvolve} and \ref{fig:twoSiteEvolveErr}). In addition to the error from neglecting or truncating the Schrieffer-Wolff transformation, there is the error from truncating the effective Hamiltonian \eqref{eq:HamEffH1spin2}. This accumulates with time and is in principle of order $J^2 t$, but may also just result in a sort of rescaling of the time axis. Also the local observables considered relax relatively quickly, making this second source of error less important.

In the remainder of the article, the initial state \eqref{eq:NeelState}, evolved with the Heisenberg Hamiltonian \eqref{eq:HamEffH1spin} will be called $\phi(t)$. If it is evolved with the Hubbard Hamiltonian \eqref{eq:Ham-Hubbard}, it will be called $\tilde\phi(t)$.
\begin{equation}\label{eq:evolvedStates}
 |\phi(t)\ket:=e^{\hat H_\eff t/i\hbar}|\phi\ket\quad \text{and}\quad
|\tilde\phi(t)\ket:=e^{\hat H t/i\hbar}|\phi\ket,
\end{equation}
and concerning observables we have explained that
\begin{equation} \label{eq:observable}
\bra \hat O\ket_{\tilde\phi} = \bra\hat O\ket_\phi + \mc O(\SW^2).
\end{equation}

To illustrate the considerations above, let us shortly regard the case of an isolated double well (two sites). Hamiltonian and $\SW$ read (cf.\ Appendix~\ref{sec:derivation})
\begin{align*}
 \hat H &= \hat H_t + \hat H_0\\
\hat H_0 &= (U+\Delta)\kzD\bzD+(U-\Delta)\kDz\bDz\\
\hat H_t &= -\t\big(\kzD\bud + \kDz\bdu+h.c.\big)\\
\SW &=\textstyle \/{i\t}{U-\Delta}\kDz\big(\bdu+\bud\big)\\
&\phantom{=}\textstyle - \/{i\t}{U+\Delta}\kzD\big(\bdu+\bud\big) +h.c.\,.
\end{align*}
With this, the effective Hamiltonian and the transformed initial state are
\begin{align}
 \hat H_{\eff} &=\textstyle -\/{4\t^2 U}{U^2-\Delta^2}\hat{\vec{S}}_1\*\hat{\vec{S}}_2 +\mc O(\SW^4)\nonumber\\
e^{-i\SW}\kud&=(1-i\SW)\kud+\mc O(\SW^2)\nonumber\\ \label{eq:2site-psi}
& \simeq\textstyle \kud +\/{\t}{U-\Delta} \kDz-\/{\t}{U+\Delta} \kzD.
\end{align}
So a magnetic state with one particle per site corresponds in the experimentally realized  Hubbard model to the magnetic state plus an admixture of states with doubly occupied and empty sites, Fig.~\ref{fig:lattice_DoubleWell_SW-Trafo}. The original Hamiltonian generates with $\hat H_\t$ doubly occupied sites to first order in $\t$. Conversely, in the (full) effective model, such terms are at least of order $\t^2$ (in the two site case here, actually of order $\t^4$).
\begin{figure}[ht]
\epsfig{file=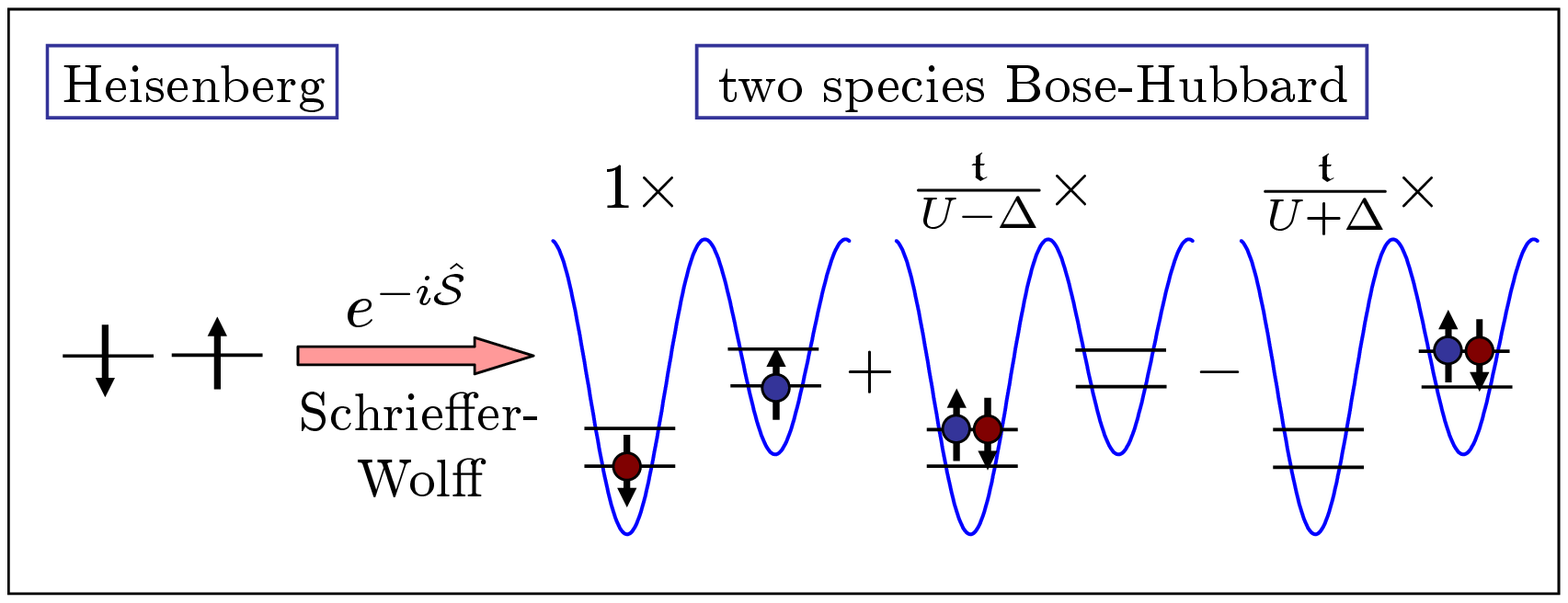, angle=0,width=1\linewidth}
\caption{\label{fig:lattice_DoubleWell_SW-Trafo}
Spin up (down) states of the effective magnetic model are not to be identified directly with a boson of species $\ua$ (species $\da$) in the experimentally realized Bose-Hubbard model. A spin up in the effective model corresponds rather to a boson of species $\ua$ with a cloud of hole-double-occupancy fluctuations $a_{\sigma i}\to e^{i\SW} a_{\sigma i} e^{-i\SW}$. In the vicinity of $\Delta=U$, the correspondence breaks down.}
\end{figure}

\subsection{Symmetry between the ferromagnetic and the antiferromagnetic cases}  \label{sec:symmetry}
The Néel state $|\phi\ket$, \eqref{eq:NeelState}, and the effective Hamiltonian \eqref{eq:HamEffH1spin} are both real in the $\{S^z_i\}_i$-eigenbasis $\mc B:=\{|\vec{\sigma}\ket=|\sigma_1\sigma_2\dots\ket\}$ (real coefficients and matrix elements). Typical observables $\hat O$ of interest like $\hat S^z_i$ for the magnetization or $\hat S^z_i\hat S^z_j$ and $\hat S^+_i\hat S^-_j+\hat S^+_j\hat S^-_i$ for correlators are real in that basis and selfadjoint. It follows that the corresponding expectation values $\bra\hat O \ket_{\phi(t)}$ are identical for the Heisenberg ferromagnet ($J=1$)  and antiferromagnet ($J=-1$): Let $o_{\vec{\sigma},\vec{\sigma}'}:=\bra\vec{\sigma}|\hat O|\vec{\sigma}'\ket$, $\phi_{\vec{\sigma}}:=\bra\vec{\sigma}|\phi\ket$, and $u_{\operatorname{(a)fm},\vec{\sigma},\vec{\sigma}'}(t):=\bra\vec{\sigma}|\hat U_{\operatorname{(a)fm}}(t)|\vec{\sigma}'\ket$ for the time evolution operator of the (anti)ferromagnetic Heisenberg model. Then
\begin{multline}
\mathbb{R}\ni 
\bra\phi|U^\dag_{\operatorname{fm}}(t)\*\hat O\* U^\pdag_{\operatorname{fm}}(t)|\phi\ket
= \left(\vec{\phi}^\dag u^\dag_{\operatorname{fm}}(t)\* o\* u^\pdag_{\operatorname{fm}}(t) \vec{\phi}\right)^*\\
= \vec{\phi}^\dag (u^\dag_{\operatorname{fm}}(t))^*\* o\* (u^\pdag_{\operatorname{fm}}(t))^* \vec{\phi}\\
= \bra\phi|U^\dag_{\operatorname{afm}}(t)\*\hat O\* U^\pdag_{\operatorname{afm}}(t)|\phi\ket.
\end{multline}

The evolution of the corresponding observable under the full Hubbard Hamiltonian $\bra\hat O \ket_{\tilde\phi(t)}$ will obey this symmetry to zeroth order in $\SW$. Typically, the resulting curve will coincide well with the corresponding Heisenberg curve. The smaller $|U^2-\Delta^2|$ is chosen, the worse the effective model will capture the actual dynamics and the stronger deviations from the corresponding Heisenberg results will be. The specific form of the deviations, however, will depend on the choice of $U$, $\Delta$, and $\t$. In particular they show no symmetry when switching between the antiferromagnetic and the ferromagnetic regimes ($\Delta\gtrless U$). To illustrate this further, several plots contain the two curves $\Delta=0$ and $\Delta=\sqrt{2}U$ which have according to \eqref{eq:HamEffH1spinCouplings} the same effective spin spin interaction strength $J$, except for the opposite sign (FM, AFM, respectively).

\subsection{Numerical method and parameters} \label{sec:numerics}
For the numerical simulation, a Krylov subspace variant \cite{Park1986-85,Hochbruck1997-34} of the time-dependent DMRG algorithm was used \cite{White2004,Daley2004,Schollwoeck2005}. For the Hubbard model, the site basis was restricted to a maximum of two particles for each species. Insensitivity of observables to the chosen maximum number of bosons per site was affirmed.
We chose lattice sizes of $L=33$ for the Bose-Hubbard model and $L=65$ sites for the Heisenberg model. Odd numbers are useful here to have reflection symmetric states. 

As a matter of fact, boundary effects are much less problematic here than in groundstate calculations, as the initial state is a product state and correlations between sites are generated inside a causal cone (the analogon of a light cone; see Section~\ref{sec:correlations}, in particular Figs.~\ref{fig:HeisenbergAntiFerroCorrConeLog} and~\ref{fig:AFM_AverCorrHubHeisenLogU8D16}). So as long as measurements are done in the middle of the system, outside of the causal cones starting from the boundary sites, results are except for exponentially small contributions identical to those of an infinite system (thermodynamic limit).

In the time evolution, the absolute difference per physical time unit between exactly evolved state and the state evolved via DMRG $||\psi_{\ud t}^{\operatorname{exact}} -\psi_{\ud t}^{\operatorname{DMRG}}||/\ud t N$ was bounded from above by $\veps=10^{-4}$ to $\veps =10^{-6}$ and the time step chosen appropriately between $\ud t=0.1$ and $0.01$. The errors were determined in a rigorous fashion, by calculating the exact value of $||\, |k+1\ket - \hat H|k\ket\, ||$, where $|k\ket$ are the Krylov vectors.
For all calculated observables, convergence in the error bound and $\ud t$ was checked. The resulting number of basis states, used to represent the time-evolved state, was $\lesssim 3000$.

\subsection{Site magnetization}  \label{sec:magnetization}
\begin{figure}[ht]
\epsfig{file=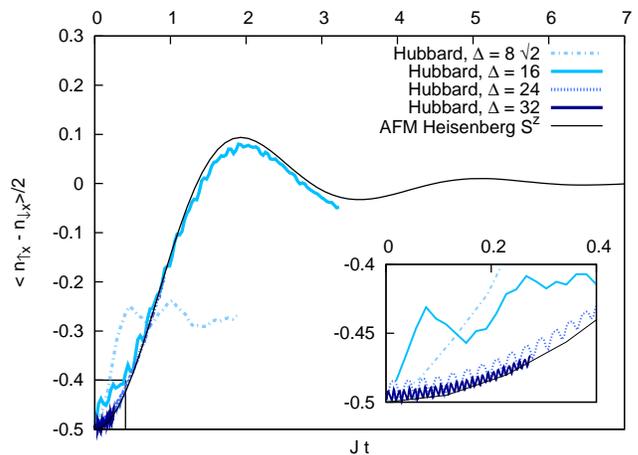, angle=0,width=1\linewidth}
\caption{\label{fig:Abweichung_Hubbard_AFM_Heisenberg_Long_T}
Evolution of the magnetization on a particular site $x$, starting from the N\'eel state and evolving with respect to the full Hubbard Hamiltonian with $U=8$ and $\Delta>U$, and the isotropic Heisenberg antiferromagnet, respectively. The first order processes occur on the time scale $\t=1$ (here $1/J$ due to the rescaling of the time axis, where time is given in units of the effective coupling $J$) and their amplitude decreases quickly with increasing $|U^2-\Delta^2|$.}
\end{figure}

\begin{figure}[ht]
\epsfig{file=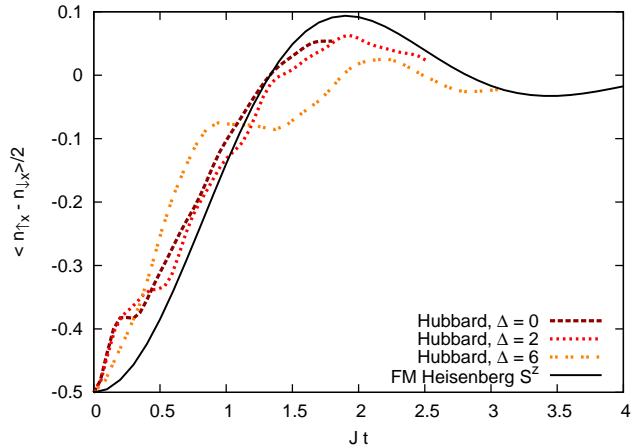, angle=0,width=1\linewidth}
\caption{\label{fig:Abweichung_Hubbard_FM_Heisenberg_Long_T}
Evolution of the magnetization on a particular site $x$, starting from the N\'eel state and evolving with respect to the full Hubbard Hamiltonian with $U=8$ and $\Delta<U$, and the isotropic Heisenberg ferromagnet, respectively. The first order processes occur on the time scale $\t=1$. Here the contributions of the first order processes cannot be made arbitrarily small as we are limited by $|U^2-\Delta^2|\leq U^2$. The Heisenberg curve here is identical to the one of the antiferromagnet in Fig.~\ref{fig:Abweichung_Hubbard_AFM_Heisenberg_Long_T} due to symmetry, see Section~\ref{sec:symmetry}. The effective coupling $J$ has the same modulus for $\Delta=0$ and $\Delta=\sqrt{2}U$, namely $|J|=4\t^2/U$, but opposite sign. The two curves show quite different behavior. There is no particular symmetry except the one for the second order physics as discussed in Section~\ref{sec:symmetry}.}
\end{figure}

\begin{figure}[ht]
\epsfig{file=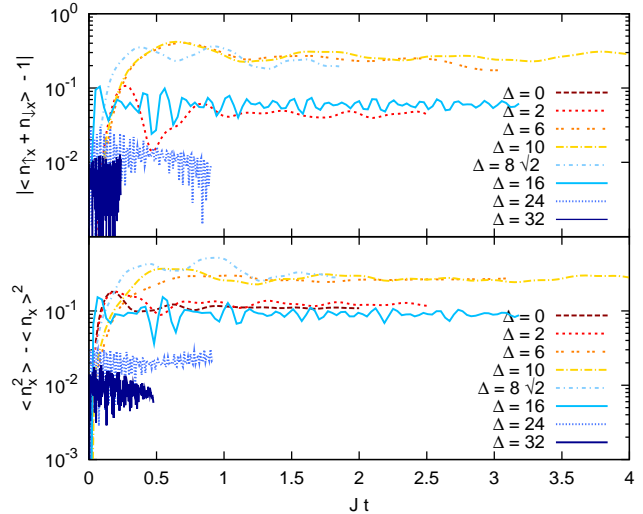, angle=0,width=1\linewidth}
\caption{\label{fig:AbweichungVonEinfachbesetzung}
Evolution of the occupation number $\bra n_{\ua x}+n_{\da x}\ket-1$ (upper panel) and its variance $\bra n_{x}^2\ket-\bra n_{x}\ket^2$ (lower panel) on a particular site $x$, starting from the N\'eel state and evolving with respect to the full Hubbard Hamiltonian with $U=8$ and several $\Delta$. The two quantities should be exactly zero, if the analogy to the spin model was exact. The analogy breaks when $|U^2-\Delta^2|$ goes to zero. In the special case $\Delta=0$, the system is (additionally to the invariance under translations by two sites) invariant under translation by one site plus interchange of particle species. Hence $\bra n_{\ua x}+n_{\da x}\ket=1\,\forall_t$ for $\Delta=0$.}
\end{figure}

Figures~\ref{fig:Abweichung_Hubbard_AFM_Heisenberg_Long_T}--\ref{fig:AbweichungVonEinfachbesetzung} show the evolution of the site magnetization $m_x=\bra\hat S^z_x\ket_{\phi}$ in the Heisenberg model and the corresponding quantity $\tilde m_x=\bra n_{\ua x}-n_{\da x}\ket_{\tilde\phi}/2$ ($=\bra\hat S^z_x\ket_{\phi}+\mc O(\SW^2)$ according to Section~\ref{sec:errors}) for the full Hubbard Hamiltonian. For the latter, times were rescaled by the coupling constant $J$, \eqref{eq:HamEffH1spinCouplings}, of the corresponding effective spin model. Site $x$ was chosen to be in the middle of the system in order to avoid finite size effects (cf.\ Section~\ref{sec:numerics}). For the Heisenberg model (in the thermodynamic limit), the site magnetization obeys for symmetry reasons $m_{x+1}=-m_x$. Analogously, due to invariance under translations by an even number of sites and particle number conservation, one has for the Hubbard model (again in the thermodynamic limit) $\bra n_{\sigma x}+n_{\sigma x+1}\ket_{\tilde\phi}=1\,\forall_t$ and hence $\tilde m_{x+1}=-\tilde m_{x}$, and $\bra n_{\ua x+1}+n_{\da x+1}\ket_{\tilde\phi}=2-\bra n_{\ua x}+n_{\da x}\ket_{\tilde\phi}$ for all times. As discussed in Section~\ref{sec:numerics}, deviations of our numerical results from the thermodynamic limit are negligible, although the simulations are carried out with finite lattice sizes.

The larger $|U^2-\Delta^2|$ is (for fixed $\t=1$), the better the curves for the full Hubbard Hamiltonian coincide with those of the Heisenberg model. This is consistent with Sec.~\ref{sec:effectiveModel} as the perturbative derivation of the effective model becomes exact in this limit.
Note that the deviations between the measurements stem from two contributions here: (a) failure of preparing the correct $\mc H_1^\orig$ state, i.e.\ applying the Schrieffer-Wolff transformation at $t=0$, (b) failure of measuring $\hat S^z_x$ instead of $e^{-i\SW}\hat S^z_x e^{i\SW}=\hat S^z_x-i[\SW,\hat S^z_x]+\mc O(\SW^2)$. The weight of those errors which are of order $\SW^2$ vanishes only far away from $|\Delta|=U$.

For the Heisenberg model, we observe relaxation of the site magnetization from $\pm 1/2$ to $0$. The oscillations of this observable occur on the time scale $1/J$. The relaxation is possible due to the continuous spectrum of the Heisenberg model (in the thermodynamic limit). Here, the convergence to a steady state is connected to a phase averaging effect, as is typical for integrable systems. Analytically this can be seen in a time-dependent mean field treatment of the Heisenberg model which we present in Section~\ref{sec:relaxation:HeisenbMeanField}. For the staggered magnetization ($m_x$) one obtains (in this approximation) a damped oscillation with the amplitude decaying as $\sim 1/t^{3/2}$. This coincides well with the DMRG data, giving support to the mean field approach; see Section~\ref{sec:relaxation:HeisenbMeanField}, Fig.~\ref{fig:Heisenb_MeanField_evolveNeel}.

For large $|U^2-\Delta^2|$ the Hubbard dynamics clearly follow the curves obtained with the Heisenberg model (second order processes); Fig.~\ref{fig:Abweichung_Hubbard_AFM_Heisenberg_Long_T} and \ref{fig:Abweichung_Hubbard_FM_Heisenberg_Long_T}. On the shorter time-scale $1/\t=1$ ($J/\t$ in the rescaled plots), corresponding to first order processes, small oscillations around the Heisenberg curves are visible. Their amplitude decreases with increasing $|U^2-\Delta^2|$. 
The perturbative treatment of the system, leading to the isotropic Heisenberg model, breaks down for $|\Delta|\sim U$. In this case, the two boson species cannot be interpreted as spin up or down states anymore and one has an appreciable amount of double occupancies in the system as demonstrated in Fig.~\ref{fig:AbweichungVonEinfachbesetzung}.

\begin{figure}[ht]
\epsfig{file=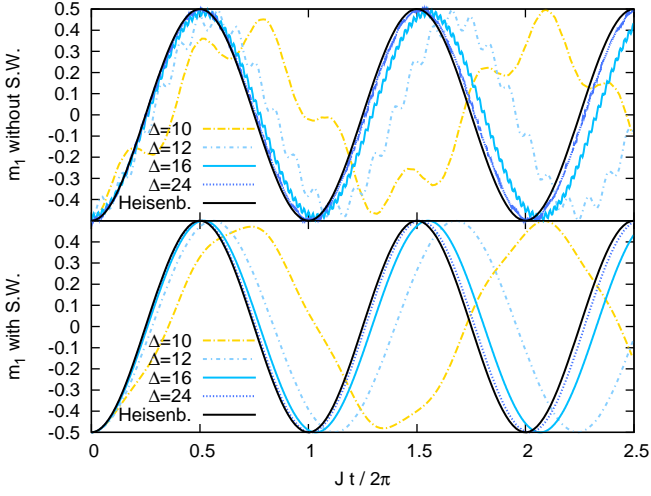, angle=0,width=1\linewidth}
\caption{\label{fig:twoSiteEvolve}
Evolution of the magnetization ($\hat {\mathfrak m}_1:=(n_{\ua 1}-n_{\da 1})/2$) on one site of an isolated double well, starting from the N\'eel state and evolving with respect to the full two-site Hubbard Hamiltonian with $U=8$ and $\Delta=10,16,24,32$, and the isotropic Heisenberg antiferromagnet, respectively. Contrary to the case of an infinite lattice, the magnetization does not relax here.
The upper panel shows $\bra \hat {\mathfrak m}_1\ket_{\tilde \phi}\equiv\bra\phi|e^{-\hat Ht/i\hbar}\hat {\mathfrak m}_1 e^{\hat Ht/i\hbar}|\phi\ket$. In the lower panel shows $\bra\phi|e^{i\hat S}e^{-\hat Ht/i\hbar}e^{-i\hat S}\hat {\mathfrak m}_1 e^{i\hat S}e^{\hat Ht/i\hbar}e^{-i\hat S}|\phi\ket$, i.e.\ there the Schrieffer-Wolff transformation was accounted for ($\SW$ correct to $\mc O(\/{\t}{U\pm\Delta})$). As discussed in Section~\ref{sec:errors}, the stretching in the curves w.r.t.\ time results from terms of order $J^2$ in the effective Hamiltonian. They originate from fourth-order hopping processes.}
\end{figure}

Finally, we want to compare those results to the dynamics for isolated double wells as addressed experimentally in \cite{Foelling2007,Trotzky2008-319}. Fig.~\ref{fig:twoSiteEvolve} shows for this case the dynamics of the site magnetization again for the Hubbard model at various $\Delta>U$ and the corresponding antiferromagnetic Heisenberg model. The decisive difference is that no equilibration is possible in this case. This is due to the fact that the Hamiltonian has only a few discrete eigenvalues here, as opposed to a gapless continuous spectrum for the lattice systems in the thermodynamic limit. In the two-site Heisenberg model we have only two states in the basis of the $S^z=0$ Hilbert space. The two eigenstates have energy difference $J$.
The magnetization curve for the Heisenberg curve is hence just a cosine with frequency $J$ and constant amplitude $1$. The dynamics of the corresponding two-site Hubbard model is determined by three (discrete) incommensurate frequencies. The magnetization is hence not completely periodic, but due to the relation to Heisenberg model, a frequency $\sim J$ is still dominating. No sign of relaxation is visible.

\begin{figure}[ht]
\epsfig{file=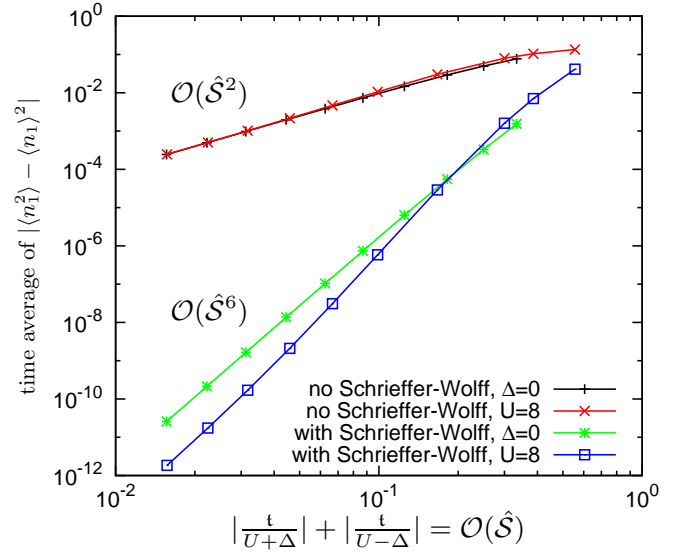, angle=0,width=1\linewidth}
\caption{\label{fig:twoSiteEvolveErr}
Time average of the particle number variance on one site of an isolated double well ($n_1\equiv n_{\ua 1}+n_{\da 1}$), evolving with respect to the full two-site Hubbard Hamiltonian with several $U$ and $\Delta$ ($J=4\t^2 U/(U^2-\Delta^2)$). The variance should be exactly zero, if the analogy to the spin model was exact as we would have exactly one particle per site then. As discussed in Section~\ref{sec:errors}, the error is of $\mc O(\SW^2)$, if the Schrieffer-Wolff transformation is neglected completely, \eqref{eq:observable}, and of $\mc O(\SW^4)$, if its first order approximation \eqref{eq:SWtrafo-result} is used. In the special case of the isolated double well, the second order terms in $\SW$ vanish (because $\hat H^0_t\equiv 0$ here, see \eqref{eq:Ham-Ht0}). Hence, we actually observe $\mc O(\SW^6)$ instead of $\mc O(\SW^4)$. The quantity on the $x$-axis quantifies $\mc O(\SW)$. For each curve, either $\Delta$ or $U$ was kept constant and the other parameter varied. Compare also to Figs.~\ref{fig:Abweichung_Hubbard_AFM_Heisenberg_Long_T} and \ref{fig:AbweichungVonEinfachbesetzung}.}
\end{figure}

 As mentioned above, the differences between Hubbard and Heisenberg dynamics stem from the fact that the two Schrieffer-Wolff transformations in \eqref{sec:errors} have been neglected. Fig.~\ref{fig:twoSiteEvolve} shows in the lower panel the site magnetization for the case where both error sources (a) and (b) have been corrected. Although this should be hard to implement experimentally, it is unproblematic for our numerical analysis. We apply the Schrieffer-Wolff transformation \eqref{eq:SWtrafo-result}, correct up to $\mc O(\/{\t}{U\pm\Delta})$, to the initial state before the Hubbard time evolution and its inverse before the measurement. As discussed in Section~\ref{sec:errors}, the remaining deviations from the Heisenberg curve are then only of order $\SW^4$; Figs.~\ref{fig:twoSiteEvolve} and \ref{fig:twoSiteEvolveErr}.

\subsection{Correlation functions}  \label{sec:correlations}
\begin{figure}[ht]
\epsfig{file=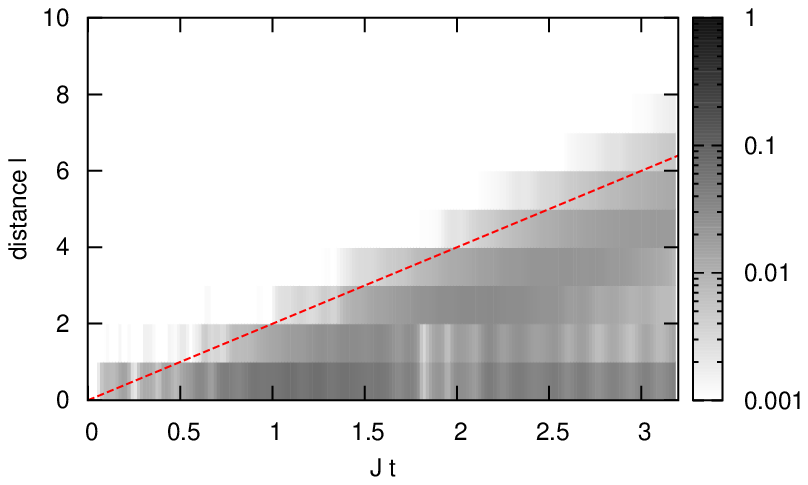, angle=0,width=1\linewidth}
\caption{\label{fig:AFM_CorrConeLogU8D16}
Evolution of the analogon $\/{1}{4}\bra (n_{\ua x}-n_{\da x})(n_{\ua x+\ell}-n_{\da x+\ell})\ket_{\tilde\phi}-\/{1}{4}\bra n_{\ua x}-n_{\da x}\ket_{\tilde\phi}\bra n_{\ua x+\ell}-n_{\da x+\ell}\ket_{\tilde\phi}$ of the mag\-ne\-ti\-za\-tion-magnetization correlation function, starting from the N\'eel state and evolving with respect to the full Hubbard Hamiltonian with $U=8$ and $\Delta=16$. The plot shows the absolute value of the correlator in logarithmic scaling. The line denotes the maximum group velocity $2J$ of the Heisenberg model.}
\end{figure}

\begin{figure}[ht]
\epsfig{file=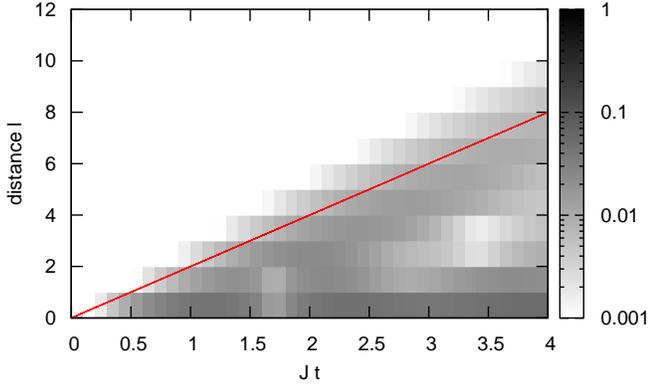, angle=0,width=1\linewidth}
\caption{\label{fig:HeisenbergAntiFerroCorrConeLog}
Evolution of the magnetization-magnetization correlation function $\bra\hat S^z_x\hat S^z_{x+\ell}\ket_{\phi}-\bra\hat S^z_x\ket_{\phi}\bra\hat S^z_{x+\ell}\ket_{\phi}$, starting from the N\'eel state and evolving with respect to the isotropic Heisenberg antiferromagnet. The plot shows the absolute value of the correlator in logarithmic scaling. The line denotes the maximum group velocity $2J$ of the Heisenberg model.}
\end{figure}

\begin{figure}[ht]
\epsfig{file=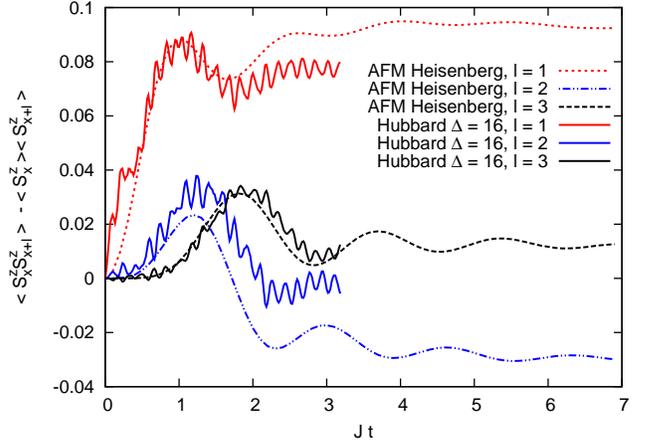, angle=0,width=1\linewidth}
\caption{\label{fig:NextNeighTwoPointSzInTime_Hubbard_Heisenberg}
Evolution of short range magnetization-magnetization correlation function (distances $\ell=1,2,3$), starting from the N\'eel state and evolving with respect to the full Hubbard Hamiltonian with $U=8$ and $\Delta=16$ and the isotropic Heisenberg antiferromagnet, respectively. With increasing time, deviations between Heisenberg and Hubbard dynamics become more pronounced than for the magnetization in Fig.~\ref{fig:Abweichung_Hubbard_AFM_Heisenberg_Long_T}. However, in both cases tendency toward equilibration to a steady state is visible.}
\end{figure}

\begin{figure}[ht]
\epsfig{file=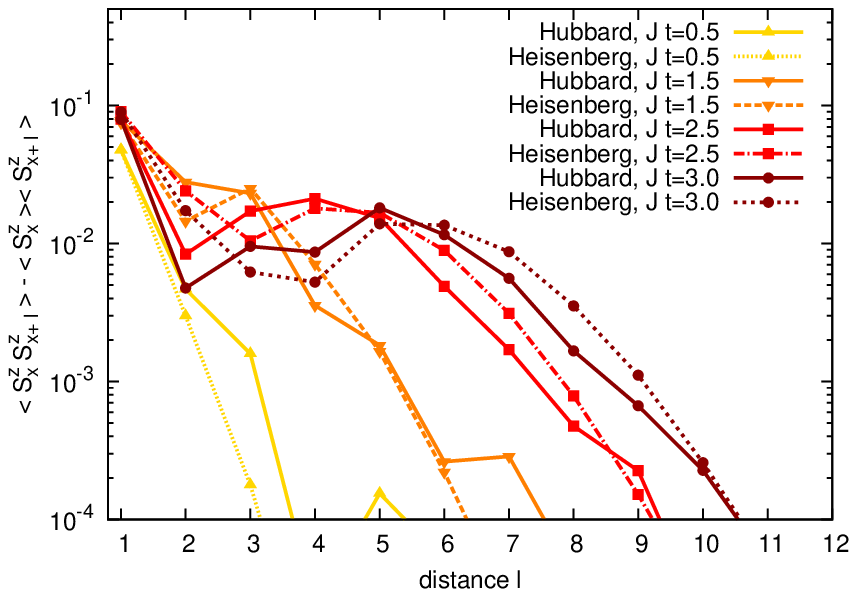, angle=0,width=1\linewidth}
\caption{\label{fig:AFM_AverCorrHubHeisenLogU8D16}
Evolution of the magnetization-magnetization correlation function, starting from the N\'eel state and evolving with respect to the full Hubbard Hamiltonian with $U=8$ and $\Delta=16$ and the isotropic Heisenberg antiferromagnet, respectively. See also Fig.~\ref{fig:NextNeighTwoPointSzInTime_Hubbard_Heisenberg}.}
\end{figure}

The correlation functions in Figs.~\ref{fig:AFM_CorrConeLogU8D16}--\ref{fig:AFM_AverCorrHubHeisenLogU8D16} support on the one hand the results already obtained from the magnetization dynamics in Section~\ref{sec:magnetization}. On the other hand one also sees here explicitly that correlations spread out inside a causal cone (analogon of a light cone) defined by the maximum group velocity. The latter coincides for large $|U^2-\Delta^2|$ with the maximum group velocity $2J$ of the Heisenberg model. One also notes here that equilibration to a steady state occurs first for small subsystems. This issue will be discussed in Section~\ref{sec:relaxation}.

\subsection{Momentum distribution and correlators}
Experimental access to onsite magnetization or, correspondingly, the particle number difference (``spin imbalance'') has already been demonstrated \cite{Foelling2007,Trotzky2008-319}. However there is no direct access to the real-space correlators. As it turns out, the standard experimental observable for experiments with ultracold atoms, the momentum distribution $\bra n_k\ket=\bra n_{\ua k}+ n_{\da k}\ket$, is to zeroth order in $\SW$ constant in time. 
It measures to this order simply the particle density which is in the limit of Heisenberg dynamics $\Delta\gg U$ very close to one.
\begin{equation}
n_{\vec{k}} 
= \/{1}{N}\sum_{i} n_i+\/{1}{N}\sum_{\sigma, i\neq j} e^{i\vec{k}\*(\vec{r}_i-\vec{r}_j)} a_{\sigma i}^\dag a_{\sigma j}^\pdag
\end{equation}
It follows with \eqref{eq:observable}
\begin{equation}
\label{eq:momentumDensity}
\bra n_{\vec{k}} \ket_{\tilde\phi}
= \bra n_{\vec{k}} \ket_\phi + \mc O(\SW^2) = 1 + \mc O(\SW^2).
\end{equation}

Hence one needs to go beyond the measurement of the momentum distribution.
By analysis of shot-noise of absorption images taken by time-of-flight measurements, one obtains momentum-space particle density correlation functions \cite{Altman2004-70,Foelling2005-434}. Experimentally available are $\bra n_{\sigma \vec{k}} n_{\sigma\vec{k}'}\ket$ and $\bra n_{ \vec{k}} n_{\vec{k}'}\ket=\sum_{\sigma\sigma'}\bra n_{\sigma \vec{k}} n_{\sigma'\vec{k}'}\ket$  and hence also $\sum_\sigma \bra n_{\sigma \vec{k}} n_{-\sigma\vec{k}'}\ket$. In the following, we will again use the approximation $\bra \hat O\ket_{\tilde\phi} = \bra\hat O\ket_\phi + \mc O(\SW^2) \simeq \bra\hat O\ket_\phi$.
\begin{multline}
\label{eq:momentumDensityDensity}
\bra n_{\sigma \vec{k}} n_{\sigma'\vec{k}'}\ket_{\tilde\phi} = \/{1}{N^2}\sum_{ij,nm} e^{i\vec{k}\*(\vec{r}_i-\vec{r}_j)}e^{i\vec{k}'\*(\vec{r}_m-\vec{r}_n)}\\
\bra a_{\sigma i}^\dag a_{\sigma j}^\pdag a_{\sigma' m}^\dag a_{\sigma' n}^\pdag\ket_{\tilde\phi}
\end{multline}
Those give information about long-range spin correlations.
\begin{align} 
\bra a_{\ua i}^\dag a_{\ua j}^\pdag a_{\ua m}^\dag a_{\ua n}^\pdag\ket_{\tilde\phi}
\simeq& \bra \delta_{ij}\delta_{mn}n_{\ua i}n_{\ua m}\nonumber\\
& + (1 - \delta_{ij})\delta_{in}\delta_{jm} a_{\ua i}^\dag a_{\ua j}^\pdag a_{\ua j}^\dag a_{\ua i}^\pdag\ket_{\phi}\nonumber\\
=&\textstyle \bra \delta_{ij}\delta_{mn}(\/{1}{2}+\hat S_i^z)(\/{1}{2}+\hat S_m^z)\nonumber\\
&\textstyle + (1-\delta_{ij})\delta_{in}\delta_{jm}(\/{1}{2}+\hat S_i^z)(\/{3}{2}+\hat S_j^z)
\ket_{\phi}
\label{eq:spaceDensityDensityUpUp}
\end{align}
for the observable $\bra n_{\ua \vec{k}} n_{\ua\vec{k}'}\ket$ and
\begin{align}
\textstyle\sum_\sigma&\bra a_{\sigma i}^\dag a_{\sigma j}^\pdag a_{-\sigma m}^\dag a_{-\sigma n}^\pdag\ket_{\tilde\phi}\nonumber\\
\simeq&\textstyle \bra \delta_{ij}\delta_{mn}\sum_\sigma n_{\sigma i}n_{-\sigma m}\nonumber\\
&\textstyle + (1 - \delta_{ij})\delta_{in}\delta_{jm} \sum_\sigma a_{\sigma i}^\dag a_{\sigma j}^\pdag a_{-\sigma j}^\dag a_{-\sigma i}^\pdag\ket_{\phi}\nonumber\\
=&\textstyle \bra \delta_{ij}\delta_{mn}(\/{1}{2}-2\hat S_i^z\hat S_m^z)\nonumber\\
&\textstyle + (1-\delta_{ij})\delta_{in}\delta_{jm}2(\hat S_i^x\hat S_m^x+\hat S_i^y\hat S_m^y)
\ket_{\phi}
\label{eq:spaceDensityDensitySmS}
\end{align}
for the observable $\sum_\sigma \bra n_{\sigma \vec{k}} n_{-\sigma\vec{k}'}\ket$.
This also reflects the fact that to zeroth order of $\SW$, there are no double occupancies with respect to the original basis. However, there is an admixture of them, contributing in second order  of $\SW$, $\bra n_{\ua i}n_{\da i}\ket_{\tilde\phi}=\mc O(\SW^2)$.
Inserting \eqref{eq:spaceDensityDensityUpUp} and \eqref{eq:spaceDensityDensitySmS} to \eqref{eq:momentumDensityDensity} yields
\begin{align}
\bra  n_{\ua\vec{k}}n_{\ua\vec{k}'}  \ket_{\tilde\phi}
&\simeq \textstyle \/{1}{4} - \/{1}{N} + \/{3}{4}\delta_{\vec{k}\vec{k}'} +\bra\hat S^z_{\Delta\vec{k}}\hat S^z_{-\Delta\vec{k}}\ket_\phi\nonumber\\
&=\bra  n_{\da\vec{k}}n_{\da\vec{k}'}  \ket_{\phi}
\label{eq:momentumDensityDensityUpUp}
\end{align}
and
\begin{multline}
\sum_\sigma \bra  n_{\sigma\vec{k}}n_{-\sigma\vec{k}'} \ket_{\tilde\phi}\\
\simeq \textstyle \/{1}{2}  - \/{1}{N} + 2\bra\hat S^x_{\Delta\vec{k}}\hat S^x_{-\Delta\vec{k}}+\hat S^y_{\Delta\vec{k}}\hat S^y_{-\Delta\vec{k}}\ket_\phi
\end{multline}
where $\Delta \vec{k}\equiv\vec{k}-\vec{k}'$, $\hat S_{\vec{q}}^\alpha\equiv \/{1}{N}\sum_i e^{i\vec{q}\*\vec{r}_i}\hat S_i^\alpha$, $\hat S^\alpha\equiv\/{1}{N}\sum_i \hat S_i^\alpha$, and $\hat S^z|\phi(t)\ket=0$ were used.

\begin{figure}[ht]
\epsfig{file=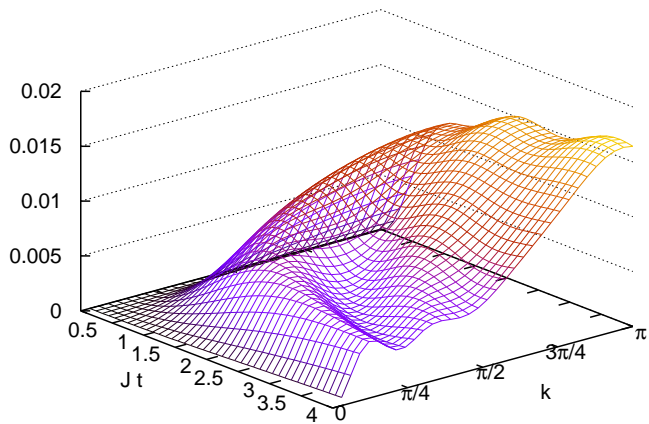, angle=0,width=1\linewidth}
\caption{\label{fig:SqSq-HeisenbergAFM} Evolution of the momentum-space spin-spin correlator $\bra\hat S^z_{\Delta k}\hat S^z_{-\Delta k}\ket_\phi$ for the Heisenberg antiferromagnet. The correlator corresponds according to Eq.~\eqref{eq:momentumDensityDensityUpUp} to the density-density correlator in the Hubbard model and is available in experiments with ultracold atoms \cite{Altman2004-70}. The initial state \eqref{eq:NeelState} is uncorrelated. Correlations build up on the time scale $1/J$. Finite-size effects have been corrected (see text).}
\end{figure}

\begin{figure}[ht]
\epsfig{file=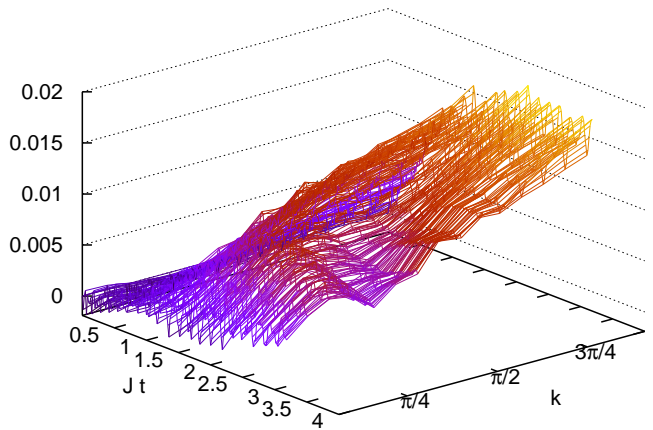, angle=0,width=1\linewidth}
\caption{\label{fig:SqSq-HubbardD16}  Evolution of the momentum space density-density correlator $\/{1}{N_k}\sum_{k}\bra n_{\ua k+\Delta k}n_{\ua k}\ket_{\tilde\phi}$ (minus the trivial parts on the right hand side of Eq.~\eqref{eq:momentumDensityDensityUpUp}) in the Hubbard model with $U=8$ and $\Delta=16$. Except for quick oscillations on the time scale $t=1/\t$, the result reflects the evolution of the corresponding spin-spin correlator in the Heisenberg model, Fig.~\ref{fig:SqSq-HeisenbergAFM}. Finite-size effects have been corrected and first order hopping contributions entering through the nearest neighbor correlator $\bra a^\dag_i a_j\ket_{\tilde\phi}$ were removed (see text).}
\end{figure}

\begin{figure}[ht]
\epsfig{file=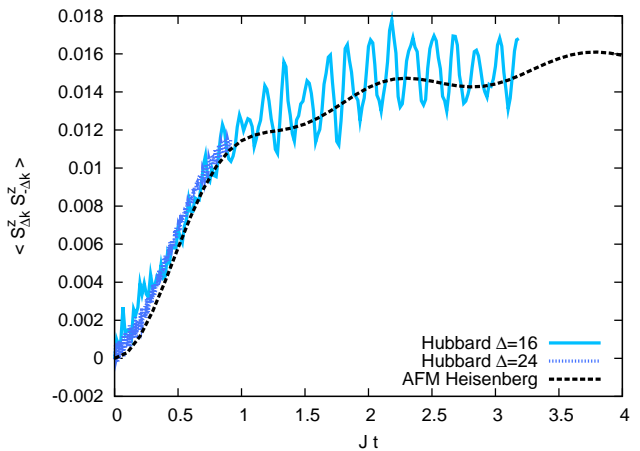, angle=0,width=1\linewidth}
\caption{\label{fig:SqSq-HubbardD16-piHalf}  Evolution of the momentum space density-density correlator $\/{1}{N_k}\sum_{k}\bra n_{\ua k+\pi}n_{\ua k}\ket_{\tilde\phi}$ for $\Delta k=\pi$ (minus the trivial parts on the right hand side of Eq.~\eqref{eq:momentumDensityDensityUpUp}) in the Hubbard model with $U=8$ and $\Delta=16,24$. 
The Hubbard results follow once more the Heisenberg curves, except for some quick oscillations due to first order hopping processes which die out for $|\Delta|$ far from $U$.}
\end{figure}

\begin{figure}[ht]
\epsfig{file=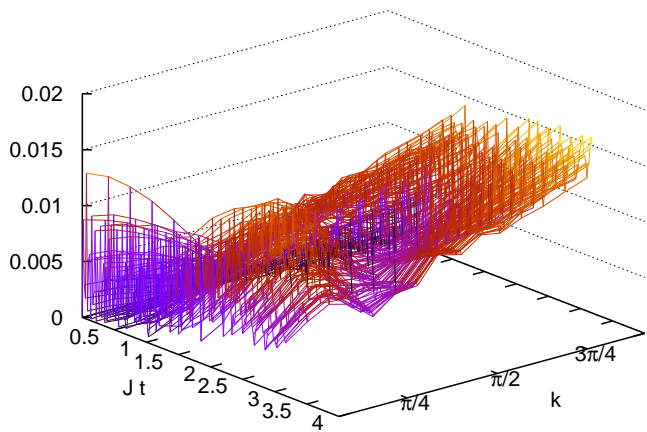, angle=0,width=1\linewidth}
\caption{\label{fig:SqSq-HubbardD16_noCorrection}  The same observable as in Fig.~\ref{fig:SqSq-HubbardD16} except, that the observable has \emph{not} been corrected for finite size effects and the effects of the correlator $\bra a^\dag_i a_j\ket_{\tilde\phi}$. We see here clearly, that those corrections of the raw data are important to achieve comparability to the corresponding Heisenberg result in Fig.~\ref{fig:SqSq-HeisenbergAFM}.}
\end{figure}

A numerical comparison of the evolution of the momentum-space spin-spin (density-density) correlator for the Heisenberg and the Hubbard models is given in Figs.~\ref{fig:SqSq-HeisenbergAFM}-\ref{fig:SqSq-HubbardD16-piHalf}.
To achieve such a good agreement, two corrections were necessary that are described in more detail in Appendix~\ref{sec:postprocess-nk-nk}. First of all one needs to correct for finite size effects. Secondly, single particle Green's functions $\bra a^\dag_i a^\pdag_j\ket$ enter which are trivial when evolving with the Heisenberg model ($\bra a^\dag_i a^\pdag_j\ket_{\phi}=\delta_{ij} n_{\ua i}(t)$), but have contributions of $\mc O(\SW^2)$, when evolving with the Hubbard Hamiltonian. In the comparison of both evolutions, they can hence be understood as a major carrier of disturbance, reflecting first order processes in the Hubbard model. To achieve comparability it would be desirable to remove contributions from $\bra a^\dag_i a^\pdag_j\ket$ completely. This would be possible for our numerical analysis. In a corresponding experiment however, the quantities are not available. Hence we confined ourselves to removing only the contributions from nearest neighbor correlators $\bra a^\dag_i a^\pdag_{i\pm 1}\ket$. As Figures~\ref{fig:SqSq-HeisenbergAFM}-\ref{fig:SqSq-HubbardD16-piHalf} demonstrate that this is already sufficient and Fig.~\ref{fig:SqSq-HubbardD16_noCorrection} that it is necessary. The experimental observation of the nearest-neighbor correlators is within reach \cite{Flesch2008-78}.

The specific form of the momentum-space correlation function can be understood with the causal cone behavior of the corresponding real-space correlators discussed in Section~\ref{sec:correlations}. At the beginning of time evolution, correlations for small distances build up (e.g.\ due to the spin flip terms $\hat S_i^+\hat S_{i\pm 1}^-$ in the Heisenberg model). This corresponds in the momentum space representation to correlations for large $\Delta k$. As the correlations spread out in real-space, correlations for smaller momenta $\Delta k$ build up.

\section{Relaxation to steady states} \label{sec:relaxation}

\subsection{General features} \label{sec:relaxation:general}
\begin{figure}[ht]
\epsfig{file=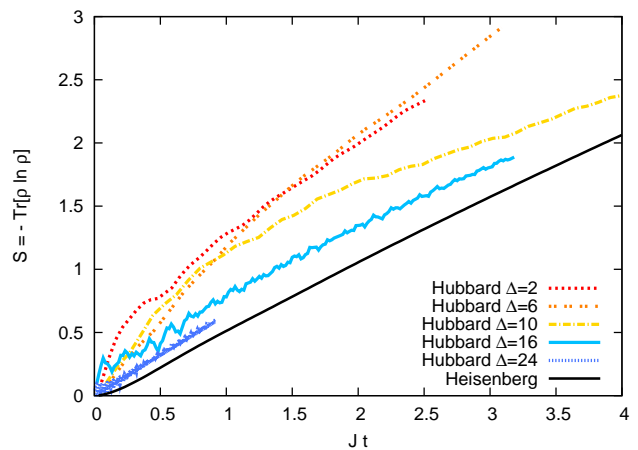, angle=0,width=1\linewidth}
\caption{\label{fig:EntanglementInTime}
For the initial state \eqref{eq:NeelState}, evolution of the entanglement entropy with respect to a partition of the system into left and right half. The growth is roughly linear in time (compare e.g.\ to \cite{Calabrese2005}) resulting in an exponential increase in the computation time required for the simulation. The more important first order processes are, the faster the entanglement entropy increases. The tuning from the Heisenberg model ($|U^2-\Delta^2|\to\infty$), where no first order processes occur, to the regime $|\Delta|\sim U=8$ can be understood as a smooth increase in the number of relevant degrees of freedom, resulting in a stronger entanglement growth. The entanglement entropies for the Heisenberg ferro- and antiferromagnet are identical, because the corresponding density matrices are in the $\{\hat S^z_i\}$-eigenbasis simply related by complex conjugation; cf.\ Section~\ref{sec:symmetry}.}
\end{figure}

Contrary to the setup of isolated double wells, one observes for the many-particle dynamics in our setup a relaxation for local quantities. This may be seen as an indicator for convergence of the states of subsystems with finite real-space extent to a steady state. 
Recently, the mechanism of how such a relaxation may occur was clarified for (free) integrable systems \cite{Barthel2008-100}. Corresponding examples can also be found in \cite{Rigol2007-98,Cazalilla2006-97,Cramer2007,Gangardt2007}.
The setup considered in this paper could be used to study experimentally such relaxation processes -- in particular, the differences for the nonintegrable Bose-Hubbard model and the Bethe ansatz integrable Heisenberg model. Experimental investigations would be very useful here, as the  fast entanglement growth during time evolution, Fig.~\ref{fig:EntanglementInTime}, prohibits numerical access to long times and for Bethe ansatz integrable systems, analytical results are also relatively limited for such purposes.
 
Integrable many-particle systems do not relax to the well known canonical, or ``thermal'', ensembles (a fact that was already observed experimentally e.g.\ in \cite{Kinoshita2006-440}). If they relax the steady state is due to the integrals of motion to a much more constrained ensemble \cite{Rigol2007-98,Barthel2008-100}. This could be detected experimentally by comparing the steady state correlation functions after time evolution to those obtained for the corresponding thermal ensemble. The temperature should be chosen such as to have the same energy in both states. For (free) integrable models, the relaxation occurs due to a phase averaging (``dephasing'') effect \cite{Barthel2008-100}. In Section~\ref{sec:relaxation:HeisenbMeanField}, the relaxation in the Bethe ansatz integrable Heisenberg model is treated within a mean field approximation. Also in this case, relaxation is connected to a phase averaging effect.

Nonintegrable systems are generally believed to relax to a thermal ensemble due to effective scattering processes. Recent numerical analysis of such systems \cite{Manmana2007-98,Kollath2007-98,Cramer2008-101} is not yet fully conclusive due to limitations on maximum observation times (density-matrix renormalization-group) or system size (exact diagonalization). Analytical approaches are usually restricted to rather exotic models or limiting cases. See e.g.\ \cite{Eckstein2007,Moeckel2008-100} for investigations by dynamical mean-field theory (DMFT).

In our setup, the nonintegrable two-species Bose-Hubbard model could be tuned so close to the Heisenberg regime (large $|U^2-\Delta^2|$) that thermalization occurs very slowly. One might hence observe first a relaxation to a nonthermal (almost) steady state due to the integrable Heisenberg dynamics, which would then be followed by slower thermalization due to the remaining nonintegrable first order processes of the full Bose-Hubbard Hamiltonian.

\subsection{Relaxation for the Heisenberg magnet in mean field approximation} \label{sec:relaxation:HeisenbMeanField}
In this section, we investigate analytically the relaxation of the Heisenberg magnet with the initial state being the Néel state \eqref{eq:NeelState}. In particular we will derive that the (staggered) magnetization decays as $1/t^{3/2}$ due to a phase averaging effect.

The model is Bethe ansatz integrable \cite{Bethe1931,Zachary1996}. However, it is in general not possible to solve the equations of motion for arbitrary initial states. With appreciable numerical effort this has been achieved recently (only) for the initial state being the groundstate plus a one-particle excitation \cite{Caux2005-95}. To investigate the dynamics nevertheless, we hence employ a mean field approximation for the $\hat S^z_i\hat S^z_{i+1}$ term.
\begin{alignat}{2}
\hat H 
&\textstyle=  
\sum_i \big( \/{1}{2}(\hat S^+_i\hat S^-_{i+1}+\hat S^-_i\hat S^+_{i+1})
+ \hat S^z_i\hat S^z_{i+1}\big) \nonumber\\
&\textstyle\to 
\sum_i \big( \/{1}{2}(\hat S^+_i\hat S^-_{i+1}+\hat S^-_i\hat S^+_{i+1})
- 2 (-1)^i\rho_\pi(t)\hat S^z_i\big)
\end{alignat}
where the order parameter $\rho_\pi$ is the staggered magnetization
\begin{equation} \textstyle
\rho_\pi \equiv \/{1}{N}\sum_x (-1)^x \bra \hat S^z_x \ket =  \/{1}{N}\sum_x (-1)^x \bra n_x -\/{1}{2} \ket.
\end{equation}
After a Fourier and a Jordan-Wigner transformation \cite{Jordan1928,Lieb1961} with $c_i:=(-1)^{\sum_{n=1}^{i-1}(\hat S_n^z+\/{1}{2})}\hat S_i^-$ and $\hat S^z_i=c^\dag_i c^\pdag_i-\/{1}{2}$, the mean field Hamiltonian and the staggered magnetization read with $\veps_k:=\cos k$
\begin{gather} \label{eq:MF_Ham}
\hat H(t) =  
\sum_{-\/{\pi}{2}\leq k<\/{\pi}{2}} ( \veps_k c^\dag_k c^\pdag_k -2\rho_\pi(t) c^\dag_{k+\pi} c^\pdag_k),\\
 \rho_\pi(t)=\/{1}{N}\sum_{-\/{\pi}{2}\leq k<\/{\pi}{2}}2\Re\bra c^\dag_k c^\pdag_{k+\pi} \ket.
\end{gather}
The initial state is the N\'eel state \eqref{eq:NeelState} and reads in the fermionic operators for $t=0$ with $u_k(0)=v_k(0)=1/\sqrt{2}$
\begin{equation} \label{eq:MFstate}
|\phi(t)\ket = \prod_{-\/{\pi}{2}\leq k<\/{\pi}{2}}(u_k(t)c^\dag_k + v_k(t)c^\dag_{k+\pi})|0\ket.
\end{equation}
So each mode $c_k$ is in the initial state only correlated with mode $c_{k+\pi}$. As the mean field Hamiltonian \eqref{eq:MF_Ham} couples for every $k$ also just those two modes, the state remains in the form \eqref{eq:MFstate} for all times. With $i\hbar\partial_t c_k(t)=[c_k,\hat H(t)]$ one obtains the equations of motion ($\hbar=1$)
\begin{subequations} \label{eq:MFeqOfMotion}
\begin{alignat}{4}
i\partial_t u_k(t) &= &\veps_k\* u_k(t) - 2\rho_\pi(t)\* v_k(t),\\
i\partial_t v_k(t) &= -&\veps_k\* v_k(t) - 2\rho_\pi(t)\* u_k(t),
\end{alignat}
\end{subequations}
a system of $N$ coupled nonlinear differential equations.

\begin{figure}[ht]
\epsfig{file=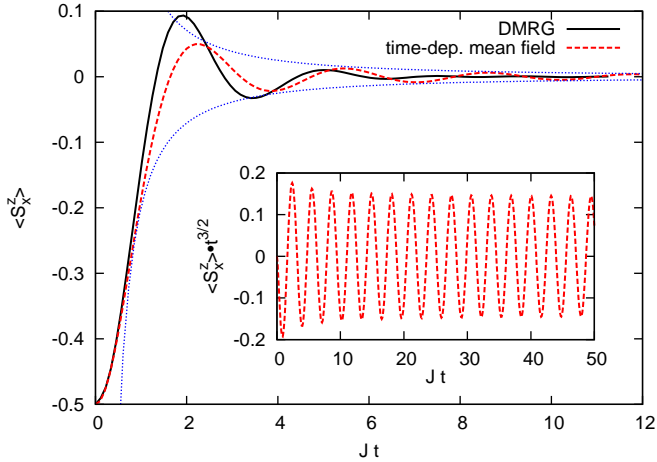, angle=0,width=1\linewidth}
\caption{\label{fig:Heisenb_MeanField_evolveNeel}
Evolution of the magnetization on a particular site $x$, starting from the N\'eel state and evolving with respect to the isotropic Heisenberg Hamiltonian, once with DMRG and once in the mean field approximation. The (staggered) magnetization shows a $1/t^{3/2}$ decay (blue). In the mean field approach one sees that local relaxation is connected to a phase averaging effect as is typical for integrable models~\cite{Barthel2008-100}; see text.}
\end{figure}

Those can be integrated numerically, yielding for the staggered magnetization $\rho_\pi(t)$ a damped oscillation decaying as $1/t^{3/2}$. Fig.~\ref{fig:Heisenb_MeanField_evolveNeel} compares $\rho_\pi(t)$ from the mean field analysis to the corresponding DMRG result (Fig.~\ref{fig:Abweichung_Hubbard_AFM_Heisenberg_Long_T}) and shows good qualitative agreement.

As demonstrated in \cite{Hastings2008}, where the same equations of motion were obtained for the evolution of a system of spinless fermions, Eq.~\eqref{eq:MFeqOfMotion} is equivalent to the equations of motion of the classical Hamiltonian 
\begin{equation} \label{eq:MFclassicalSpinHam}
 H_S = -\sum_{-{\pi}\leq k<{\pi}}2\veps_kS^z_k + \/{2}{N}\sum_{k,k'}(S^x_kS^x_{k'}+S^y_kS^y_{k'})
\end{equation}
with the Anderson pseudospin variables $S^+_k=v_k^* u_k$, $S^z_k=\/{1}{2}(|v_k|^2-|u_k|^2)$ for $-\/{\pi}{2}\leq k<\/{\pi}{2}$ and $S^x_{k+\pi}=S^x_{k}$, $S^{y,z}_{k+\pi}=-S^{y,z}_{k}$, \cite{Anderson1958-112,Hastings2008}. This Hamiltonian occurred in the mean field analysis of quenches in fermionic condensates; see e.g.\ \cite{Warner2005-71,Szymanska2005-94,Yuzbashyan2005-38,Yuzbashyan2005-72}.
From this, it is known that \eqref{eq:MFclassicalSpinHam} and hence \eqref{eq:MFeqOfMotion} are integrable \cite{Yuzbashyan2005-38} due to the $N/2$ integrals of motion $L_k^2$,
\begin{equation}
 \vec{L}_k\equiv\vec{e}_z+2\sum_{k\neq k'}\/{\vec{S}_{k'}}{\veps_k-\veps_{k'}},\quad \partial_t L_k^2=0.
\end{equation}

One can now argue that the $x$ and $y$ components of the vectors $\vec{L}_k$ will vanish for large times, as done in \cite{Hastings2008}. From this one can determine the (nonthermal) steady state, by equating $({L}^z_k(t\to\infty))^2$ with ${L}^2_k(t=0)$. The result is $\lim_{t\to\infty} S^z_k(t)=\/{1}{2} \cos k$. With $S^z_k=\/{1}{2}(|v_k|^2-|u_k|^2)$ and $|v_k|^2+|u_k|^2=1$ $\forall_t$, it follows that
\begin{gather}
 \lim_{t\to\infty} |u_k(t)| = \sqrt{1-\cos k}/\sqrt{2},\\
\lim_{t\to\infty} |v_k(t)| = \sqrt{1+\cos k}/\sqrt{2}.
\end{gather}
With the knowledge of the steady state, the $1/t^{3/2}$ decay of the magnetization $\rho_\pi$ can now be derived.

To this purpose let us first recall the general dephasing scenario for $d$-dimensional (free) integrable models. In \cite{Barthel2008-100} it was demonstrated that local observables $G(t)$ (i.e.~correlators) lead in general to expressions of the form 
\begin{equation}
\label{eq:dephasing-observable}\textstyle
G(t)=G_0+\int\ud^d k e^{i\vphi(\vec{k}) t} f(\vec{k}),
\end{equation}
where the amplitude $f(\vec{k})$ is determined by the chosen observable, the initial state, and the eigenbasis of the Hamiltonian. The phase function $\vphi(\vec{k})$ is determined by the spectrum of the Hamiltonian. Now, the quantity $G(t)$ relaxes to $G_0$ for large times if the phase function varies quickly enough in regions of the $\vec{k}$ space where the amplitude $f(\vec{k})$ is nonzero. Whether and how quickly an observable relaxes is in particular determined by contributions from points where $\vphi(\vec{k})$ is stationary or $f(\vec{k})$ diverges. For the paradigmatic scenario of $\varphi(\vec{k})\sim \varphi_0+|k|^\ell$, $f(\vec{k})\sim 1/k^m$ near a stationary point $\vec{k}_0=\vec{0}$, the integral in \eqref{eq:dephasing-observable} behaves as
\begin{equation}
\label{eq:dephasing-paradigm}\textstyle
e^{i\varphi_0 t}\int\ud^d k\/{1}{|\vec{k}|^m} e^{i|\vec{k}|^\ell t}
\sim \int\ud q \/{1}{q^\chi} e^{i q t},
\quad \chi=\/{m+\ell-d}{\ell}.
\end{equation}
Hence the time-dependent contribution to $G(t)$, for $t\to\infty$,
does not vanish if $\chi\geq 1$,
vanishes as $1/{t^{1-\chi}}$ if $0<\chi<1$,
and at least as $1/t$ if $\chi<0$,
\footnote{For free systems, the (Gaussian) state of any subsystem is fully characterized by its one-particle Green's function. This was exploited in \cite{Barthel2008-100} to derive conditions on the relaxation of subsystem states, based on the relaxation of the Green's function.}.

\begin{figure}[ht]
\epsfig{file=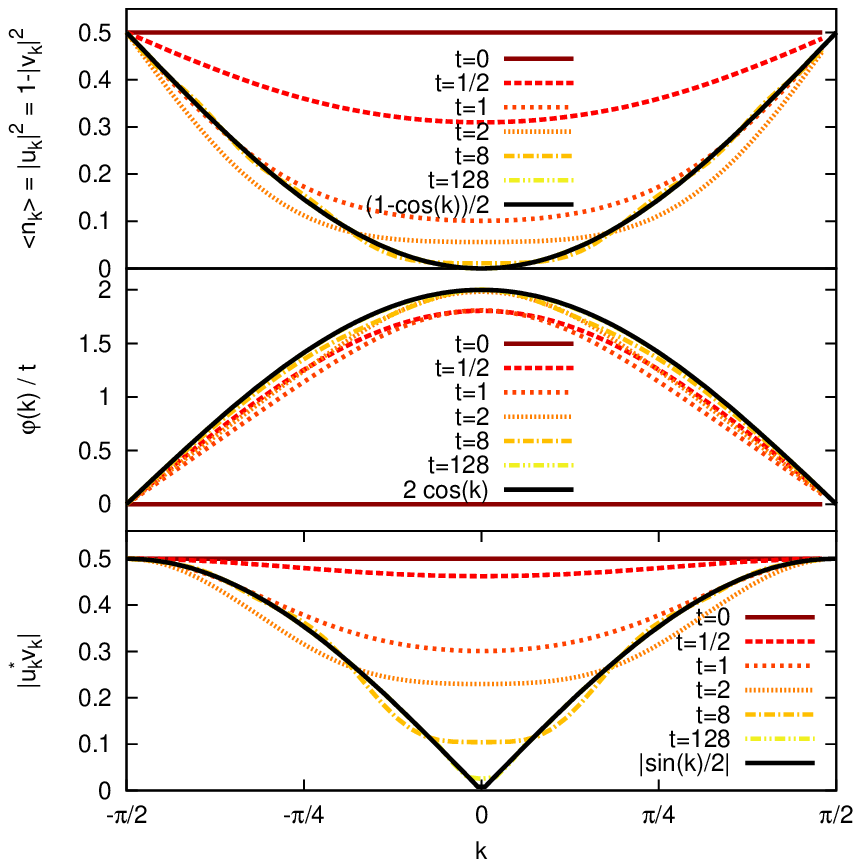, angle=0,width=1\linewidth}
\caption{\label{fig:Heisenb_MeanField_evolveNeel.t_k_nk_uvArg}
Evolution of the occupation number $\bra n_k\ket=1-\bra n_{k+\pi}\ket$, phase $\vphi(k)=\arg(u_k^*(t)v_k(t))-\arg(u_{\pi/2}^*(t)v_{\pi/2}(t))$ of the wavefunction, and $|u_k^*v_k|$ for each pair of modes $c_k$, $c_{k+\pi}$.
The initial state, the N\'eel state with $u_k=v_k=1/\sqrt{2}$, is evolved with the mean field approximation of the isotropic Heisenberg Hamiltonian \eqref{eq:MF_Ham}. Except for the lowest panel, the curves for $t=128$ coincide (within resolution of the plots) with the limiting curves  for $t\to \infty$ which are $\/{1-\cos k}{2}$, $2\cos k$, and $|\sin k|/2$, respectively.}
\end{figure}

Now we come back to the staggered magnetization. Expressed in the variables $u$ and $v$, it reads after going to the thermodynamic limit
\begin{equation}\label{eq:staggeredM_uv}
 \rho_\pi(t)=\int\ud k \Re(u_k^*v_k) = \int\ud k \Re(e^{i\vphi(k)t} f(k,t)).
\end{equation}
This is, except for the additional time dependence of the amplitude function $f(k,t)$, an integral of the form \eqref{eq:dephasing-observable}.
Presuming that $\rho_\pi$ vanishes for long times, it follows from the equations of motion \eqref{eq:MFeqOfMotion} that for large $t$, the phases of $u_k$ and $v_k$ are roughly $\pm\veps_k\*t$ and hence $\vphi(k)\approx \veps_k-\veps_{k+\pi}=2\cos k$, which is stationary (with $\ell = 2$) at $k=0$; cf.\ Fig.~\ref{fig:Heisenb_MeanField_evolveNeel.t_k_nk_uvArg}. For finite times, the amplitudes of $u_k$ and $v_k$ are finite as $|u_k|^2=\bra n_k\ket$ and $|v_k|^2=\bra n_{k+\pi}\ket=1-\bra n_k\ket$. Hence $m=0$ for $k=0$ which is also confirmed numerically in Fig.~\ref{fig:Heisenb_MeanField_evolveNeel.t_k_nk_uvArg}.
The dephasing of the staggered magnetization \eqref{eq:staggeredM_uv} is determined by the stationary point $k=0$ of $\varphi$. With $d=1$, $m=0$, and $\ell=2$ we have $\chi=\/{m+\ell-d}{\ell}=\/{1}{2}$. The phase averaging accounts hence for a factor of $1/t^{1-\chi}=1/t^{1/2}$ for the decay of the staggered magnetization. Linearizing the equations of motion around the steady state we find that $|u_0^* v_0|$ ($f(k,t)$ around $k=0$) decays as $1/\sqrt{t}$ and further that only a vicinity $|k|\lesssim 1/\sqrt{t}$ of $k=0$ is contributing to (the leading order of) the integral \eqref{eq:staggeredM_uv}. Taking all this together, we infer the $1/t^{3/2}$ decay of the staggered magnetization.

\section{Validity of the effective spin model}  \label{sec:validity}
One may be wondering why the restriction to the single-occupancy space $\mc H_1$, \eqref{eq:subspaceH1}, is justified  (if the initial state of the system is in $\mc H_1$ and we evolve with the effective Hamiltonian), although the coupling to the rest of the Hilbert space has the same strength as the coupling for dynamics inside the subspace $\mc H_1$ and although parts of the rest of the Hilbert space overlap energetically with $\mc H_1$. There could be considerable transition rates out of the subspace with (predominantly) one particle per site, rendering a description or comparison with dynamics of the effective model derived for that subspace useless. We will assess here that this is not the case (for the large-$U$ limit).

First of all, the numerical results of Section~\ref{sec:Evolution} showed that for large $|U^2-\Delta^2|$, the Hubbard curves follow quite precisely the Heisenberg curves, indicating very little transitions to other subspaces. One can also give a somewhat handwaving but rather suggestive argument.
We will show in the following that transition matrix elements leading out of $\mc H_1$ occur predominantly to states with energy difference $\sim U$, and diminish in the large-$U$ limit. Those yield therefore finite small transition amplitudes. In higher orders of the perturbation theory, there are also transitions to states with energy $\sim U\pm \Delta$, which will lead to a small (controllable) transition rate out of $\mc H_1$.

\begin{figure}[ht]
\epsfig{file=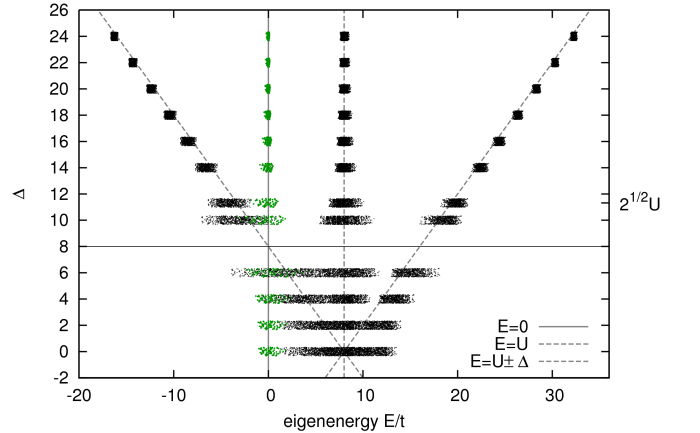, angle=0,width=1\linewidth}
\caption{\label{fig:BH2Eff_DensityOfStates}
The many-particle spectrum of the effective Hamiltonian $\hat H_\eff^0$ for the subspaces $\mc M^0$ (green) and $\mc M^1$ (black) with $U=8$ and $\Delta=2,4,6,8,10,\sqrt{2}U,14,16,\dots$ as obtained from exact diagonalization in the $S^z=0$ sector for $N=8$ sites ($\dim\mc M^0=70$, $\dim\mc M^1=2800$). Each dot corresponds to an eigenenergy. For the plot, small random numbers were added to the $\Delta$ values to give a rough impression of the density of states. The subspace $\mc M^1$ is separated energetically into $\mc M^1_{0,0}\cup \mc M^1_{0,1}$ around $E=U$ and $\mc M^1_{1,1}$, $\mc M^1_{-1,0}$ around $E=U\pm\Delta$.}
\end{figure}

In the full effective Hamiltonian \eqref{eq:HamEff}, we regard the term $\hat V=i[\SW,\hat H_\t^0]$ that generates or destroys double occupancies, i.e.\ generates transitions between subspace $\mc M^{n}$ with $n$ doubly occupied sites as a perturbation. 
\begin{equation}
 \hat H_\eff^\full = \hat H_\eff^0 + \hat V
\end{equation}
The subspaces $\mc M^n$ separate (energetically) further into $\mc M^{n}_{m-\mu,m}$ with $m$ doubly occupied and $\mu$ empty sites on sublattice $\mc A$ ($n-m$ doubly occupied and $n-\mu$ empty sites on sublattice $\mc B$); i.e.\ $\mc H_1\equiv\mc M^0_{0,0}$. 
Fig.~\ref{fig:BH2Eff_DensityOfStates} shows the many-particle spectrum of the effective Hamiltonian $\hat H_\eff^0$ for the subspaces $\mc M^0$ and $\mc M^1$ as obtained from exact diagonalization in the $S^z=0$ sector for $N=8$ sites.

The single (quasi-)particle excitations in these subspaces have energies of order $\mc O(J,\t)$ -- spinwaves and hopping of doubly occupied and empty sites. However, the subspaces overlap energetically (in the thermodynamic limit) as, in a qualitative picture, one can have $\sim N$ quasi-particle excitations resulting in the width $\sim N|J|\gg U$ of the spectrum for each subspace. Specifically for $\mc M^0$, the lower and upper bounds on the spectrum are determined by the ground state energies of the ferromagnetic and the antiferromagnetic Heisenberg models. Those are in the thermodynamic limit $E_{\operatorname{fm}}=-\/{1}{4}JN$ and $-E_{\operatorname{afm}}=(\ln 2-\/{1}{4}) JN$, \cite{Bethe1931,Medeiros1991}.

\begin{figure}[ht]
\epsfig{file=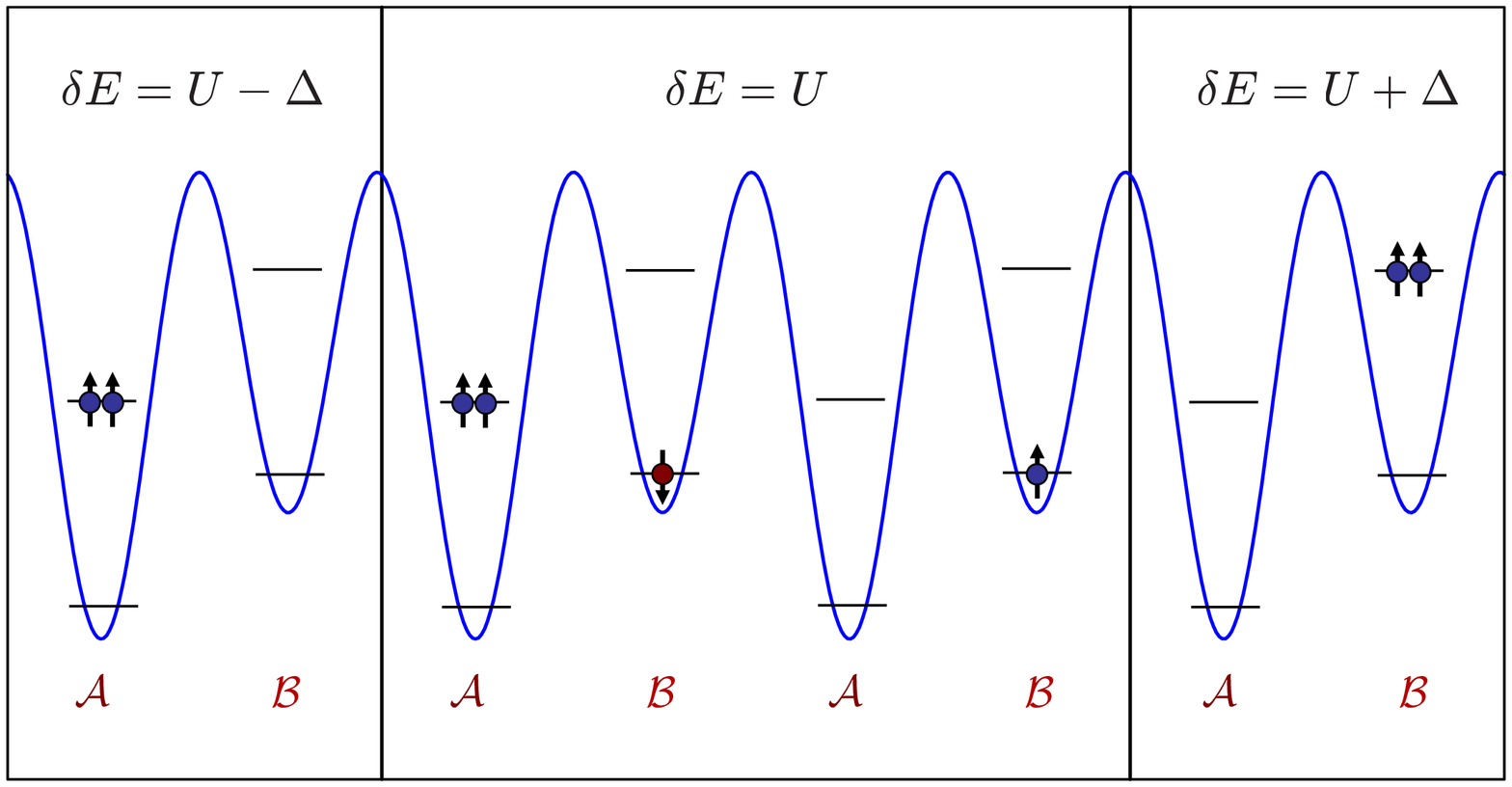, angle=0,width=1\linewidth}
\caption{\label{fig:lattice_doubleOccup_subbands}
The subspace $\mc M^1$, with exactly one doubly occupied and one empty site separates energetically into three different subbands $\mc M^1_{1,1}$, $\mc M^1_{0,0}\cup \mc M^1_{0,1}$, and $\mc M^1_{-1,0}$.
The operator $\hat V=i[\SW,\hat H_\t^0]$ maps states from $\mc M^0$ to states from $\mc M^1_{0,0}\cup \mc M^1_{0,1}$ that differ in energy by $\sim U$ (see text).}
\end{figure}

If we act on a state $|\psi\ket\in\mc M^0$ of energy $E$ with the operator $\hat V=i[\SW,\hat H_\t^0]$ (cf.\ Appendix~\ref{sec:derivation}), firstly, $\SW$ generates a double occupancy and an empty site; $|\sigma_i,\sigma_{i+1}\ket\mapsto |\sigma_i\sigma_{i+1},0\ket$ on two neighboring sites $i$ and $i+1$. Secondly, a corresponding hopping term from $\hat H_\t^0$ acts on $i$ (or $i+1$) and $i-1$ (or $i+2$) such that e.g.\ $|\sigma_{i-1},\sigma_i\sigma_{i+1},0\ket\mapsto |\sigma_{i-1}\sigma_i,\sigma_{i+1},0\ket$ or $|\sigma_i\sigma_{i+1},0,\sigma_{i+2}\ket\mapsto |\sigma_i\sigma_{i+1},\sigma_{i+2},0\ket$. Hence both, the doubly occupied and the empty site are on the same sublattice $\mc A$ or $\mc B$ and the resulting state $\hat V|\psi\ket\in\mc M^1_{0,0}\cup\mc M^1_{0,1}$ has energy $E\sim E+U$; see Fig.~\ref{fig:lattice_doubleOccup_subbands}.

Let us consider transitions from $\mc M^0$ to $\mc M^1$.
For any initial eigenstate $|i\ket\in \mc M^0$ the transition amplitude, to a state $|f\ket\in\mc M^1$ is in the Born approximation given by
\begin{equation}\label{eq:transitions}
c_{f}(T)=\/{-i}{\hbar}\int_{t_0}^T \ud t  \bra f|\hat V|i\ket e^{i\omega_{fi}t}=\mc O(\/{\t^2}{U\pm\Delta}\*\/{1}{U}).
\end{equation}
This estimate of a small (oscillating) transition amplitude follows from the consideration that nonvanishing transition elements exist only for states with energy differences $\hbar\omega_{fi}=\mc O(U)$. We have pointed out that the subspaces $\mc M^0$ and $\mc M^1$ ultimately overlap energetically. However, states $|f\ket$ and $|i\ket$ with comparable energy will have a vanishing transition matrix element: As argued above, the operator $\hat V$ generates states from $\mc M^1$ and causes a change of $\sim U$ in energy. Besides this it can create or destroy in a qualitative picture only a small number of quasi-particle excitations as it is a product of only four ladder operators. This will change the energy only by a small amount of $\mc O(J,t)$. So $\hbar\omega_{fi}$ will indeed be of order $\mc O(U)$ for all nonvanishing transition amplitudes $\bra f|\hat V|i\ket$.

\begin{figure}[ht]
\epsfig{file=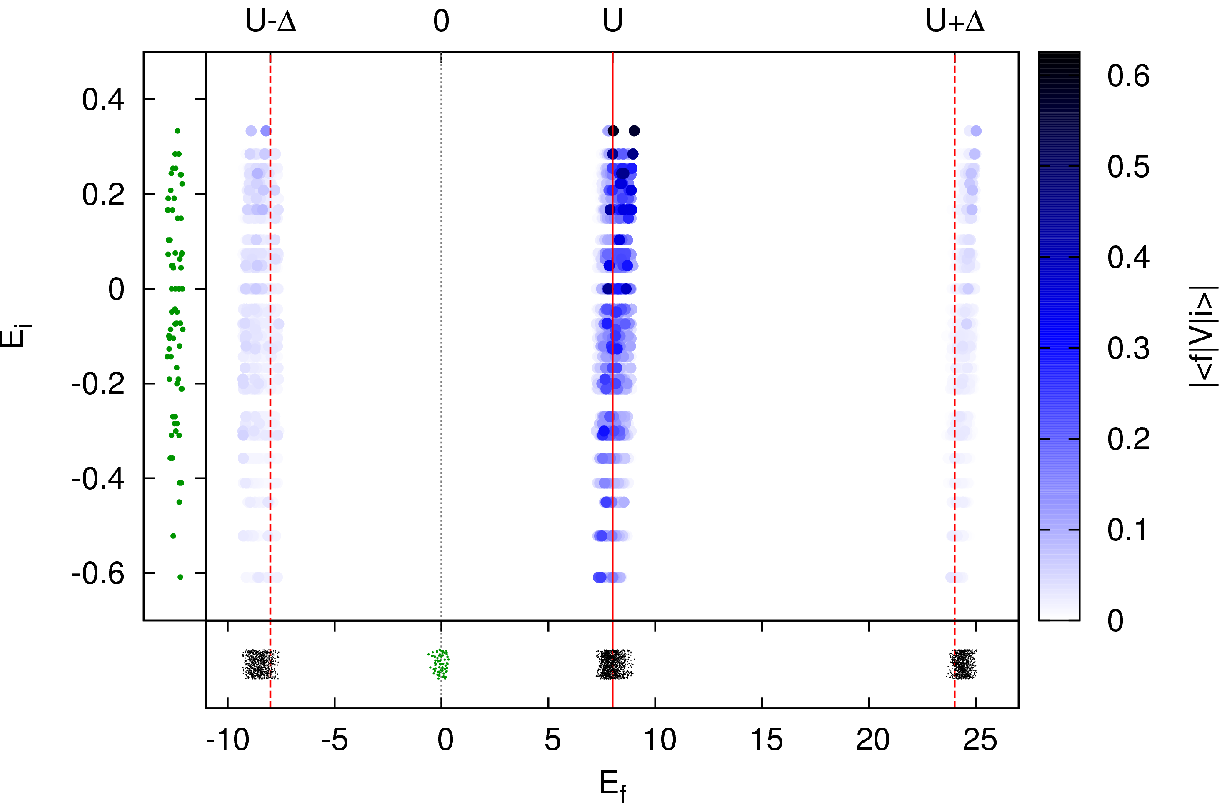, angle=0,width=1\linewidth}
\caption{\label{fig:BH2Eff_VeigenD16}
Transition matrix elements $\bra f|\hat V|i\ket$ between eigenstates of the effective Hamiltonian $\hat H_\eff^0$ for the subspaces $\mc M^0$ and $\mc M^1$ with $U=8$ and $\Delta=16$ as obtained from exact diagonalization in the $S^z=0$ sector for $N=8$ sites. Each dot corresponds to a nonzero transition matrix element. The narrow panels to the left and bottom show the corresponding eigenenergies. Nonvanishing matrix elements exist only for states with energy difference of $\mc O(U)$.}
\end{figure}

\begin{figure}[ht]
\epsfig{file=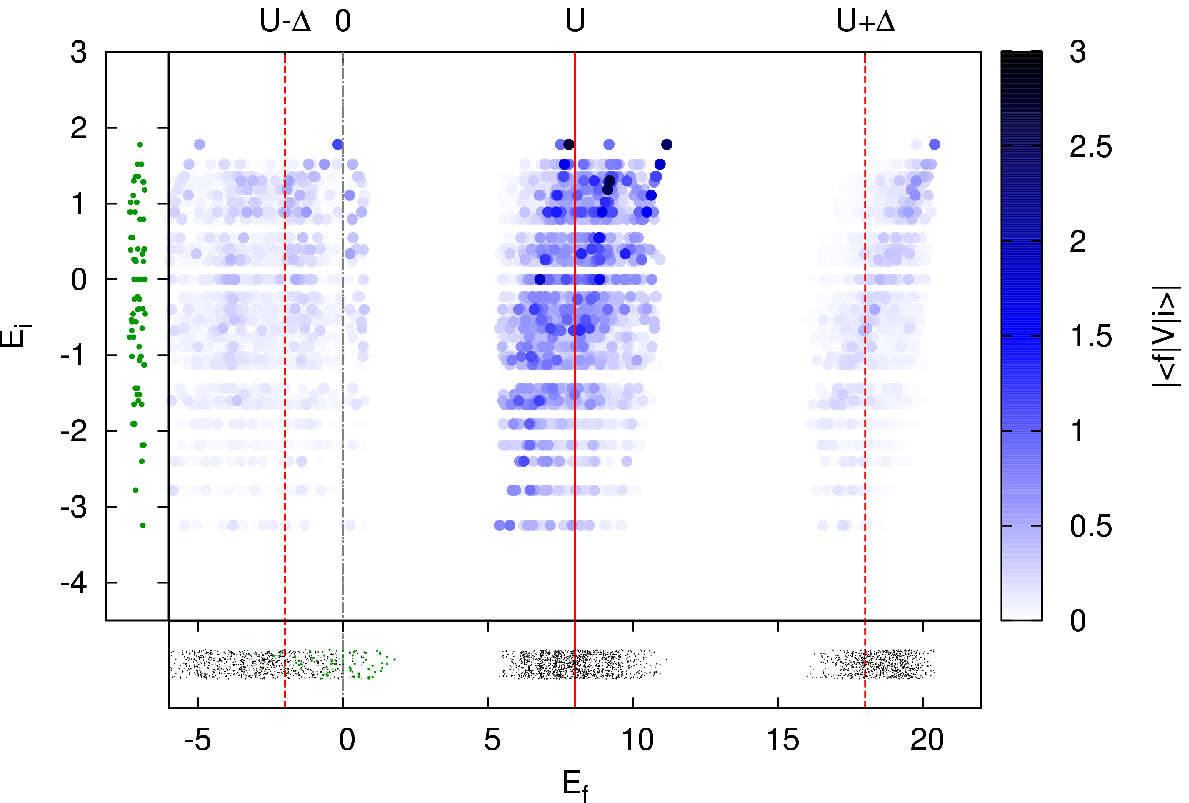, angle=0,width=1\linewidth}
\caption{\label{fig:BH2Eff_VeigenD10}
Transition matrix elements $\bra f|\hat V|i\ket$ between eigenstates of the effective Hamiltonian $\hat H_\eff^0$ for the subspaces $\mc M^0$ and $\mc M^1$ with $U=8$ and $\Delta=10$ as obtained from exact diagonalization in the $S^z=0$ sector for $N=8$ sites. Still nonvanishing matrix elements exist only for states with energy difference of $\mc O(U)$. But as $\Delta$ is closer to $U$ here, the matrix elements are larger in amplitude, \eqref{eq:transitions}, and the spectral subbands are broader due to a larger effective coupling~$J$.}
\end{figure}

To illustrate this, Figs.~\ref{fig:BH2Eff_VeigenD16} and \ref{fig:BH2Eff_VeigenD10} show the transition matrix elements $\bra f|\hat V|i\ket$ between eigenstates of the effective Hamiltonian $\hat H_\eff^0$ for the subspaces $\mc M^0$ and $\mc M^1$ as obtained from exact diagonalization in the $S^z=0$ sector for $N=8$ sites.

Small matrix elements to states with energy difference $U\pm \Delta$ remain. For unfortunate choice of $U$ and $\Delta$ one may hence encounter nonvanishing transitions to states in $\mc M^{n>0}$ with $\hbar\omega_{fi}\sim 0$. Consider e.g.\ $\Delta=2U$. In this case, two actions of the operator $\hat V$ may end up in a state $|f\ket$ with comparable energy ($U+(U-\Delta)=0$) and hence to a (finite but small) transition rate out of $\mc M^0$. By appropriate choice of the ratio $U/\Delta$, one can achieve that the effect occurs only in higher orders $\hat V$, resulting in a small transition rate. Further, the transition matrix elements itself can be made small by going to the large-$U$ limit \eqref{eq:largeU-limit}.

\section{Preparation of the antoferromagnetic groundstate by adiabatic evolution}  \label{sec:groundstates}
In the Sections~\ref{sec:Evolution} and \ref{sec:validity} we have given arguments and gathered numerical support for the fact that transition rates from the single-occupancy subspace $\mc H_1^\orig$, \eqref{eq:Schrieffer-Wolff}, to the rest of the Hilbert space can be made small for time evolution with the Hubbard Hamiltonian. If this can also be realized experimentally for sufficiently long times, it would be possible to prepare for example the ground state of the antiferromagnetic Heisenberg model by adiabatically switching on the coupling $\t'$ between initially isolated double wells, Fig.~\ref{fig:systemParameters}, i.e.\ switching from $\t'=0$ to $\t'=\t$, while $\t$ is kept constant. As demonstrated in \cite{Foelling2007,Trotzky2008-319} for the initial situation of isolated double wells, the groundstate of the single occupancy subspace \eqref{eq:subspaceH1} can be prepared experimentally.

For the adiabatic approximation \cite{Kato1950-5,Avron1987-110} to be applicable, the system needs to be gapped on the whole path in the space of system parameters except for the end point, where the gap has to close abruptly enough. As argued in Section~\ref{sec:validity}, transitions to other subspaces with (quasiparticle) double occupancies can be neglected for a certain period of time $T$ that can be made very large. So we only need to worry about transitions from the $\mc H_1^\orig$ groundstates to excited states inside the subspace, i.e.\ we need to derive conditions on the dependence of the corresponding energy gap on the hopping $\t'$ between initially isolated double wells such that $\t'=\t$ can be reached adiabatically in a finite amount of time $\tau<T$.

The quantitative condition for adiabaticity is generally stated as
\begin{equation}
\label{eq:adiabatCondition1}
\left|\/{\bra E_0(t)|\/{\ud\hat H}{\ud t}|E_n(t)\ket}{E_0(t)-E_n(t)}\right |\ll 1\quad \forall_{t\in[0,\tau],n\neq 0},
\end{equation}
where $|E_n(t)\ket$ label the energy eigenstates and $|E_0(0)\ket$ is the initial state.
Recently, substantial problems were pointed out \cite{Marzlin2006-97,Tong2005-95} and two more conditions added \cite{Tong2007-98}
\begin{gather}
\label{eq:adiabatCondition2}
\int_0^\tau\ud t\left|\/{\ud}{\ud t}\/{\bra E_0(t)|\/{\ud\hat H}{\ud t}|E_n(t)\ket}{E_0(t)-E_n(t)}\right|\ll 1,\\
\label{eq:adiabatCondition3}
\int_0^\tau\ud t\left|\/{\bra E_0(t)|\/{\ud\hat H}{\ud t}|E_n(t)\ket}{E_0(t)-E_n(t)}\right||\bra E_n(t)|\/{\ud\hat H}{\ud t}|E_m(t)\ket|\ll 1.
\end{gather}
If we have one time-dependent system parameter $p(t)$, namely the dimerization $p=\delta$, where
\begin{equation}
\delta:=\/{1-|J'/J|}{1+|J'/J|}=\/{1-|\t'/\t|^2}{1+|\t'/\t|^2},
\end{equation}
and one part of the Hamiltonian is linear in that parameter (this is the case for the effective Hamiltonian and $\delta\to 0$), the numerators of \eqref{eq:adiabatCondition1}-\eqref{eq:adiabatCondition3} are proportional to the sweeping speed $v(t):=\ud p(t)/\ud t$. The denominator is the spectral gap $E_g(t)$. Only points $p(\tau)$ in parameter space where the gap vanishes are problematic. If the gap vanishes as
\begin{equation}
E_g(t)\propto |p(\tau)-p(t)|^\nu,\quad \nu>0,
\end{equation}
$v$ should (for $t$ close to $\tau$) be reduced as $c|\tau-t|^\mu$. According to  \eqref{eq:adiabatCondition1},
\begin{equation}
1\gg  \/{c|\tau-t|^\mu}{E_g(t)}
\propto\/{|\tau-t|^\mu}{|\int_0^{\tau-t}\ud s \*s^\mu |^\nu}
\propto |\tau-t|^{\mu - \nu(\mu+1)}.
\end{equation}
Hence, only for $\nu<1$, i.e.\ for gaps that close abruptly enough, adiabaticity can be reached with $\mu \geq \/{\nu}{1-\nu}$. The second condition, \eqref{eq:adiabatCondition2}, is in this scenario fulfilled automatically, the third, \eqref{eq:adiabatCondition3}, implies $\mu >\/{\nu -1}{2-\nu}$ which is also true. 

For $J'\simeq J$, a situation which was examined intensively in the context of spin-Peierls systems, the model was first treated by Jordan-Wigner transformation and subsequent bosonization \cite{Cross1979-19}. The precise result for the excitation gap can be obtained by a mapping to the four-state Potts model \cite{Black1981-23} or conformal field theory \cite{Affleck1989-22} (see also \cite{Kumar2007-75}). The gap is given by
\begin{equation}
\label{eq:gap-rho1}
E_g(\delta)\propto \/{\delta^{2/3}}{|\ln \delta|^{1/2}}=\mc O(\delta^{2/3}).
\end{equation}

That means we have a gap with $\nu=2/3<1$ and hence  the gap can be closed in a finite amount of time with exponent $\mu=2$. This means that the dimerization $\delta$ has to be varied with speed $v(t)=c|\tau-t|^2$, hence $\delta(t) = \/{c}{2}|\tau-t|^3$. One needs thus the time $\tau=(2/c)^{1/3}$. The smaller $c$ is, the farther we are in the adiabatic regime but the longer we need for the preparation. An analysis of how small $c$ is to be chosen to achieve a given accuracy of the prepared state could be carried out along the lines of Ref.~\cite{Trebst2006-96}.

Note that in \cite{Koetsier2008-77}, it was recently discussed within a mean-field approach, how the antiferromagnetic phase of the three-dimensional Fermi-Hubbard model could be reached by adiabatic tuning of the lattice potential.

\section{Conclusion}  \label{sec:conclusion}
We have studied a setup of two species of ultracold bosonic atoms in an optical superlattice, which realizes in a certain parameter regime the Heisenberg ferro- and antiferromagnet. The focus was in particular on time evolution of nonequilibrium states. Our numerical results and analytical considerations showed that the physics of Bose-Hubbard model implemented in the experiment differs for certain parameter ranges considerably from the physics of the effective Heisenberg models. Note that this would also be true for alternative suggestions as in \cite{Kuklov2003-90,Duan2003-91,Altman2003-5,Garcia-Ripoll2004-93,Barmettler2008-78}. The spin states up and down can in general not be identified directly with a bosonic particle of one specific species. The regime where the correspondence between the two models is good, implies higher requirements on cooling and coherence (coherence time) in an experimental realization. The explicit form of the Schrieffer-Wolff transformation was used to analyze the transition rates out of the magnetic subspace of the full Hilbert space.

In contrast to the accomplished experiments \cite{Foelling2007,Trotzky2008-319} for isolated double-wells (filled each with two particles), the setup of coupled double-wells discussed here allows for relaxation of the many-particle state. In the numerics we observed indications for (local) relaxation to steady states. For the Heisenberg model in a mean field approximation, we explained  how the relaxation is connected to a phase averaging effect. This is typical for integrable models which have nonthermal steady states. Nonintegrable models are generally believed to thermalize due to effective scattering effects. Our setup can be tuned from the nonintegrable Bose-Hubbard model to the Bethe ansatz integrable Heisenberg model and could hence be used to study the differences of the relaxation processes experimentally.

Finally we argued that the groundstate of the Heisenberg antiferromagnet could be prepared by tuning an alternating hopping parameter of the superlattice adiabatically.

\acknowledgments
We thank I.\ Bloch, H.\ Capellmann, A.\ Flesch, S.\ F\"olling, and A.\ Kolezhuk for discussions. This work was supported by the DFG. T.~B. also acknowledges financial support by the Studienstiftung des Deutschen Volkes.

\appendix
\section{Derivation of the effective model by Schrieffer-Wolff transformation} \label{sec:derivation}
Here we derive the effective spin Hamiltonian \eqref{eq:HamEffH1spin} describing the physics of the two species Bose-Hubbard model \eqref{eq:Ham-Hubbard} in the subspace $\mc H_1^\orig$, \eqref{eq:Schrieffer-Wolff}, where every site is occupied by exactly one quasi-particle. The spin-spin interaction is generated by second order hopping processes of the particles. In the large-$U$ limit, transitions from $\mc H_1^\orig$ to bands with double-occupancies are energetically hindered. In particular, transitions from 
\begin{equation}
\label{eq:lowestHubbardBand}
\mc H_1=\Span\{|n_1,\dotsc,n_{N}\ket, n_{\ua i}+n_{\da i}= 1\,\,\forall_i\}
\end{equation}
to the subspace with double-occupancies (and holes), $\mc H_2$, can be treated perturbatively. 

The Hamiltonian contains terms, linear in the hopping $\t$, which couple $\mc H_1$ to the subspace with double-occupancies. We are looking for a canonical transformation $\hat H \to \hat H_{\eff}^\full:=e^{i\SW} \hat H e^{-i\SW}$ such that the single-occupancy subspace $\mc H_1$ of the resulting quasi-particles $e^{i\SW} a_{\sigma i} e^{-i\SW}$ couples only in second order to double occupancies ($\mc H_2$).

The calculation can be done in analogy to the derivation of the Kondo lattice model \cite{Kondo1964-32} from the periodic Anderson model \cite{Anderson1961-124,Schrieffer1966-149,Fazekas1999} or the $t$--$J$ model \cite{Chao1977-10} from the fermionic Hubbard model \cite{Hubbard1963-276} and is modified only by the asymmetry term $\Delta_i$ and the finite intra-species repulsion (double occupancies $|\ua\ua\ket$ and $|\da\da\ket$).

\begin{figure}[ht]
\epsfig{file=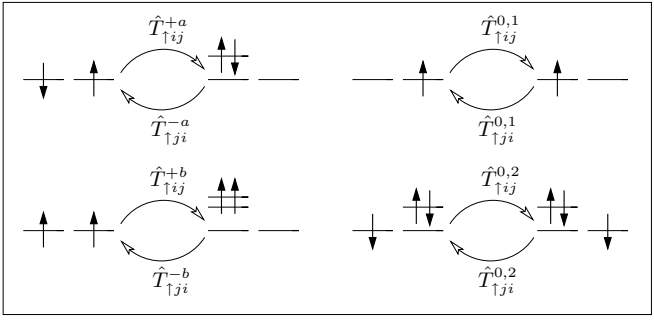, angle=0,width=1\linewidth}
\caption{Illustration of the hopping terms \eqref{eq:hopping-term-0}-\eqref{eq:hopping-term-m}. $\hat T^{\pm,\gamma}$ increases/decreases the number of doubly occupied sites and $\hat T^{0,\nu}$ leaves it unchanged.}
\label{fig:hopping-terms}
\end{figure}

Let us rewrite the Hamiltonian \eqref{eq:Ham-Hubbard}, restricted to the subspace $\mc H_1\cup\mc H_2$, in the form
\begin{align}
\label{eq:Ham-Hubbard2}
\hat H&= \hat H_0+\hat H_\t^0 + \hat H_\t^++\hat H_\t^-,\\
\hat H_0 &= \hat H_\Delta+\hat H_U\\
\hat H_\Delta &= \sum_{\sigma,i} \Delta_i n_{\sigma,i}\\
\hat H_U &=  U\sum_i n_{\ua i}n_{\da i} + \/{U_s}{2}\sum_{\sigma,i} n_{\sigma i}(n_{\sigma i}-1)\\
\hat H_\t^0 &= -\t\sum_{\sigma,\bra ij\ket,\nu} ( \hat T_{\sigma ij}^{0,\nu}+\hat T_{\sigma ji}^{0,\nu}),\quad \nu\in\{1,2\} \label{eq:Ham-Ht0}\\
\hat H_\t^\pm &= -\t\sum_{\sigma,\bra ij\ket} ( \hat T_{\sigma ij}^\pm+\hat T_{\sigma ji}^\pm),
\end{align}
where $\bra ij\ket$ runs over nearest neighbors (one index from each sublattice), $H_\t^\pm$ increases/decreases the number of doubly occupied sites and $H_\t^0$ leaves it unchanged; Fig.~\ref{fig:hopping-terms}.
\begin{alignat}{4}
\label{eq:hopping-term-0}
\hat T_{\sigma ij}^{0,1} &= \delta_{n_{i},1}&\,a_{\sigma i}^\dag a_{\sigma j}^\pdag& \,\delta_{n_{j},1},\\
\hat T_{\sigma ij}^{0,2} &= \delta_{n_{i},2}&\,a_{\sigma i}^\dag a_{\sigma j}^\pdag& \,\delta_{n_{j},2},\\
\hat T_{\sigma ij}^+ &=\delta_{n_{i},2} &\,a_{\sigma i}^\dag a_{\sigma j}^\pdag& \,\delta_{n_{j},1},\\
\hat T_{\sigma ij}^- &=\delta_{n_{i},1} &\,a_{\sigma i}^\dag a_{\sigma j}^\pdag& \,\delta_{n_{j},2},
\label{eq:hopping-term-m}
\end{alignat}
where $\delta$ denotes the Kronecker delta and in its argument, $n_i\equiv n_{\ua i}+ n_{\da i}$ denote the particle number operators.
For the operators $\hat T_{\sigma ij}^\pm$, we further distinguish between those which change the number of (a) inter-species and (b) intra-species double occupancies, see Fig.~\ref{fig:hopping-terms}. 
\begin{equation}
\hat T_{\sigma ij}^\pm=\hat T_{\sigma ij}^{\pm a}+\hat T_{\sigma ij}^{\pm b}.
\end{equation}

\subsection{Schrieffer-Wolff transformation}  \label{sec:SWtrafo}
The Hamiltonian contains terms $\hat H_\t^\pm$, linear in the hopping $\t$, which couple $\mc H_1$ to the double-occupancy subspace. We are looking for a canonical transformation $\hat H \to \hat H_{\eff}^\full=e^{i\SW} \hat H e^{-i\SW}$ such that the single-occupancy subspace $\mc H_1$ of the resulting quasi-particles $e^{i\SW} a_{\sigma i} e^{-i\SW}$ couples only in second order to double occupancies ($\mc H_2$).
\begin{align*}
\hat H_\eff^\full&=e^{i\SW} \hat H e^{-i\SW}\\
&=\hat H+i[\SW,\hat H]+\/{i^2}{2}[\SW,[\SW,\hat H]]+{\mc O}(\SW^3\hat H)\\
&=\hat H_0 +\hat H_\t^0 + \hat H_\t^++\hat H_\t^- +i[\SW,\hat H]+\/{i^2}{2}[\SW,[\SW,\hat H]]+\dots\\
\end{align*}
In the first commutator, the contribution $[\SW,\hat H_0]$ dominates and we therefore look for a generator $\SW$ such that 
\begin{equation}
\label{eq:SWtrafo}
i[\SW,\hat H_0] = -(\hat H_\t^++\hat H_\t^-).
\end{equation}
From this equation follows with $\hat H_0=\mc O(U,U_s,\Delta)$ and $\hat H_\t=\mc O(\t)$ that $\SW=\mc O(\/{\t}{(U,U_s,\Delta)})$ and hence
\begin{equation*}
\hat H_\eff^\full = \hat H_0 +\hat H_\t^0 +i[\SW,\hat H_\t]+\/{i^2}{2}[\SW,[\SW,\hat H]]+\mc O(\/{\t^3}{(U,U_s,\Delta)^2}).
\end{equation*}
This yields for \eqref{eq:SWtrafo} the solution
\begin{multline}
\label{eq:SWtrafo-result}
\SW = i\sum_{\sigma,\bra ij\ket} \big(
 \/{\t}{U+\Delta_i-\Delta_j} \hat T^{+a}_{\sigma ij} + \/{\t}{U+\Delta_j-\Delta_i} \hat T^{+a}_{\sigma ji}\\
 +\/{\t}{U_s+\Delta_i-\Delta_j} \hat T^{+b}_{\sigma ij} + \/{\t}{U_s+\Delta_j-\Delta_i} \hat T^{+b}_{\sigma ji} 
 - h.c. \big)
\end{multline}
With $\Delta_i$ from \eqref{eq:anisotropy}, we can now state more precisely
\begin{equation}
\label{eq:SWtrafo-order}
\SW=\mc O\left(\/{\t}{U \pm \Delta},\/{\t}{U_s \pm \Delta}\right).
\end{equation}
So the perturbative treatment will break down near the crossing point $\Delta=U$ and for large hopping $\t$.

\subsection{Effective spin Hamiltonian for half filling}
The full effective Hamiltonian reads
\begin{multline}
\label{eq:HamEff}
\hat H_\eff^\full = \hat H_\t^0 + \hat H_U + \hat H_\Delta\\
 + \/{i}{2}[\SW,\hat H_\t^++\hat H_\t^-] + i[\SW,\hat H_\t^0]+\mc O(\t^3).
\end{multline}
The commutator terms still couple $\mc H_1$ with the rest of the Hilbert space (subspaces with differing numbers of doubly occupied sites). However, this coupling is now not $\mc O(\t)$, as in the original Hamiltonian \eqref{eq:Ham-Hubbard2}, but of $\mc O(\t^2)$. This was achieved by the Schrieffer-Wolff transformation $\SW$, which replaces our original particles $a_{\sigma i}$, by particles with a cloud of hole-double-occupancy fluctuations $a_{\sigma i}\to e^{i\SW} a_{\sigma i} e^{-i\SW}$.
In the single-occupancy subspace $\mc H_1$ at half filling, \eqref{eq:lowestHubbardBand}, \eqref{eq:half-filling}, $\hat H_\t^0$, $\hat H_\Delta$, $\hat H_U$, $i[\SW,\hat H_\t^0]$ and the terms of third order in the hopping are all ineffective such that we are left with
\begin{equation}
\label{eq:HamEffH1}
\hat H_\eff:=\hat H_\eff^\full|_{\mc H_1} = \/{i}{2}[\SW,\hat H_\t^++\hat H_\t^-]_{\mc H_1}+\mc O(\t^4).
\end{equation}
The commutator consists of hopping terms via virtual double-occupancy states. They are of the form $\hat T^{-\gamma}_{\sigma'ji}\hat T^{+\gamma}_{\sigma ij}$ and can be rephrased as spin-spin interactions. With 
\begin{align}
\textstyle\sum_{\sigma}\hat T^{-a}_{\sigma ji}\hat T^{+a}_{\sigma ij}|_{\mc H_1} &=(1-4\hat S^z_i\hat S^z_{j})/2,\\
\textstyle\sum_{\sigma}\hat T^{-b}_{\sigma ji}\hat T^{+b}_{\sigma ij}|_{\mc H_1} &=1+4\hat S^z_i\hat S^z_{j},\\
\textstyle\sum_{\sigma}\hat T^{-a}_{-\sigma ji}\hat T^{+a}_{\sigma ij}|_{\mc H_1} &=\hat S^+_i\hat S^-_{j}+\hat S^-_i\hat S^+_{j},
\end{align}
the effective Hamiltonian \eqref{eq:HamEffH1} reads
\begin{multline}
\label{eq:HamEffH1spin2}
\hat H_\eff = 
- J \sum_{\bra ij\ket} (\hat S^x_i\hat S^x_{j}+\hat S^y_i\hat S^y_{j})\\
+ (J-J_{s}) \sum_{\bra ij\ket} \hat S^z_i\hat S^z_{j} +\mc O(\t^4),
\end{multline}
where (see also \cite{Trotzky2008-319})
\begin{gather}
\label{eq:HamEffH1spinCouplings2}
J 
=\/{4\t^2U}{U^2-\Delta^2},\quad
J_{s} 
=2\/{4\t^2U_s}{U_s^2-\Delta^2}.
\end{gather}
This Hamiltonian is, except for higher order effects, the XXZ model. In the bulk of the article we specialize to $U=U_s$, i.e.\ $J-J_s=-J$ and have hence the isotropic Heisenberg ferromagnet for $\Delta<U$ ($J>0$) and the isotropic antiferromagnet for $\Delta>U$ ($J<0$).
 As was already pointed out, the full effective Hamiltonian \eqref{eq:HamEff} still contains a coupling to the subspace with one double-occupancy (of quasi-particles). The approximation made by neglecting it is discussed in Section~\ref{sec:validity}. A peculiarity of our situation is that, due to half-filling of both particle species, we are restricted to the $S^z=0$ sector of the Heisenberg model.

One can go to higher orders in the perturbative treatment of the hopping, by adding higher order terms to the generator $\SW$ of the Schrieffer-Wolff transformation. The next order term $\SW^{(2)}$ has to be chosen such that it eliminates the term $i[\SW^{(1)},\hat H_\t^0]$ in \eqref{eq:HamEffH1}. This would result in a further contribution to the effective spin Hamiltonian, namely next nearest neighbor and four-spin interactions, generated by sequences of four virtual hopping events.

\section{Postprocessing of density-density correlators}  \label{sec:postprocess-nk-nk}
\subsection{Elimination of finite-size effects for the numerics}  \label{sec:postprocess_finite-size}
The numerics were done for a finite size system (cf.\ Section~\ref{sec:numerics}). To correct for the resulting finite-size effect is simple for the calculation of the spin-spin correlators in the Heisenberg model in the r.h.s.\ of \eqref{eq:spaceDensityDensityUpUp} and \eqref{eq:spaceDensityDensitySmS}. One can use
\begin{equation}
 \bra \hat S^\alpha_i \hat S^\beta_j\ket_\phi \mapsto  \bra \hat S^\alpha_{x_i} \hat S^\beta_{x_i+j-i}\ket_\phi,
\end{equation}
where $x_i$ is some site in the middle of the system that is odd (even) for odd (even) i. This corresponds to the invariance of the systems under translations by multiples of two sites in the thermodynamic limit.

The momentum-space density-density correlators, \eqref{eq:momentumDensityDensityUpUp}, in the Hubbard model are determined from the  real-space four point correlators \eqref{eq:spaceDensityDensityUpUp}. To eliminate finite-size effects of those, we note first that due to the restriction of all correlations to a (causal) time-space cone, the four point correlators behave for large distances as (spin indices suppressed)
\begin{align} \label{eq:fourPointCorr-longDistance}
C_{ijn}&:=\bra a_i^\dag a_j^\pdag a_{n+j-i}^\dag a_n^\pdag\ket\nonumber\\
&\to
\bra a_i^\dag a_j^\pdag\ket\bra a_{n+j-i}^\dag a_n^\pdag\ket\nonumber\\
&\phantom{\to} + [\bra a_i^\dag a_n^\pdag\ket\bra a_{n+j-i}^\dag a_j^\pdag\ket+(\delta_{i,n}-\delta_{i,j})\bra n_i\ket].
\end{align}
Consequently, the quantity
\begin{multline}
 C_{ijn}':= \bra a_i^\dag a_j^\pdag a_{n+j-i}^\dag a_n^\pdag\ket
- \bra a_i^\dag a_j^\pdag\ket\bra a_{n+j-i}^\dag a_n^\pdag\ket\\
-[\bra a_i^\dag a_n^\pdag\ket\bra a_{n+j-i}^\dag a_j^\pdag\ket+(\delta_{i,n}-\delta_{i,j})\bra n_i\ket]
\end{multline}
is localized; it has support only for $j$ and $n$ inside the causal cone centered at site $i$ (cmp.\ to Section~\ref{sec:correlations}). So the value of $C_{ijn}$ for the thermodynamic limit is approximated well by 
\begin{multline} \label{eq:fourPointCorr-thermoLimit}
 C_{ijn}'':= C_{ijn}'
+ g_{i,j}g_{n+j-i,n} \\
+ [g_{i,n}g_{n+j-i,j}+(\delta_{i,n}-\delta_{i,j})g_{i,i}],
\end{multline}
where  $g_{i,j}$ is the (approximate) single particle Green's function in the thermodynamic limit
\begin{equation}
 g_{i,j} \equiv \bra a_{x_i}^\dag a_{x_i+j-i}^\pdag\ket,
\end{equation}
and for odd (even) $i$, $x_i$ is an odd (even) site in the middle of the system.\\

\subsection{Reduction of effects from first order processes for numerics and experiments}  \label{sec:postprocess_first-order}
As disclosed by equation \eqref{eq:fourPointCorr-longDistance} or \eqref{eq:fourPointCorr-thermoLimit} the four point correlators entering the momentum-space density-density correlator contain contributions from products of single-particle correlators $\bra a_{i}^\dag a_{j}^\pdag\ket_{\tilde \phi}$.
Those are trivial when evolving with the Heisenberg model: $\bra a^\dag_i a^\pdag_j\ket_{\phi}=\delta_{ij} n_{\ua i}(t)$, as there is exactly one particle per site. But according to \eqref{eq:observable}, they have contributions of $\mc O(\SW^2)$, when evolving with the Hubbard Hamiltonian. In the comparison of observables evolved with both models, those correlators enter hence as a major carrier of disturbance. To achieve comparability it would be desirable to remove contributions from $\bra a^\dag_i a^\pdag_j\ket$ completely. This would be possible for our numerical analysis. In a corresponding experiment however, the quantities are not available. Hence we confined ourselves to removing only the contributions from nearest neighbor correlators $\bra a^\dag_i a^\pdag_{i\pm 1}\ket$. That means $g_{i,j}$ in \eqref{eq:fourPointCorr-thermoLimit} is set to zero for $j=i\pm 1$.
This was already sufficient to demonstrate the correspondence of the dynamics if we are safely in the large-$U$ limit \eqref{eq:largeU-limit}. An experimental procedure for the measurement  of the nearest-neighbor correlator was suggested in \cite{Flesch2008-78}. Hence, the same manipulations might be carried out for experimentally obtained momentum-space density-density correlators.

\bibliographystyle{prsty} 

\end{document}